\documentclass[final,12pt,onecolumn]{article}

\usepackage[dvips]{graphicx}
\usepackage{verbatim}
\usepackage{multirow}
\usepackage{url}
\usepackage[square]{natbib}

\def\gsim{\;\lower4pt\hbox{${\buildrel\displaystyle >\over\sim}$}\;}
\def\lsim{\;\lower4pt\hbox{${\buildrel\displaystyle <\over\sim}$}\;}
\def\grls{\;\lower4pt\hbox{${\buildrel\displaystyle >\over <}$}\;}

\newcommand{\ve}[1]{\mathbf{#1}}
\newcommand{\altaffilmark}[1]{$^#1$}
\newcommand\addr[2]{{\footnotesize \it $^{#1}$#2}\\}
\newcommand{\sun}{S}

\begin{document}

\title{Solar LImb Prominence CAtcher and Tracker (SLIPCAT): An Automated System and Its Preliminary Statistical Results}

\author{Yuming Wang\altaffilmark{1}, Hao Cao\altaffilmark{1}, Junhong Chen\altaffilmark{1},
Tengfei Zhang\altaffilmark{1}, Sijie Yu\altaffilmark{1},\\[1pt] 
Huinan Zheng\altaffilmark{1}, Chenglong Shen\altaffilmark{1}, Jie
Zhang\altaffilmark{2}, and S. Wang\altaffilmark{1}\\[1pt]
\addr{1}{Key Laboratory of Basic Plasma Physics of CAS,
School of Earth \& Space Sciences,}
\addr{}{University of Science \& Technology of China, Hefei, Anhui 230026, China;} 
\addr{}{Contact:
ymwang@ustc.edu.cn} \addr{2}{Department of Computational and
Data Sciences, George Mason University,}
\addr{}{4400 University Dr.,
Fairfax, VA 22030, USA}}

\maketitle
\tableofcontents

\begin{abstract}
In this paper, we present an automated system, which has the capability to catch and track solar limb prominences based on observations from EUV 304~\AA\ passband.  The characteristic parameters and their evolution, including height, position angle, area, length and brightness, are obtained without manual interventions. By applying the system to the STEREO-B/SECCHI/EUVI 304~\AA\ data during 2007 April -- 2009 October, we obtain a total of 9477 well-tracked prominences and a catalog of these events available online at \url{http://space.ustc.edu.cn/dreams/slipcat/}. A detailed analysis of these prominences suggests that the system has a rather good performance. We have obtained several interesting statistical results based on the catalog. Most prominences appear below the latitude of 60 degrees and at the height of about 26 Mm above the solar surface. Most of them are quite stable during the period they are tracked. Nevertheless, some prominences have an upward speed of more than 100 km/s, and some others show significant downward and/or azimuthal speeds. There are strong correlations among the brightness, area and height. The expansion of a prominence is probably one major cause of its fading during the rising or erupting process.
\end{abstract}

\section{Introduction}
Solar prominences, also called filaments when they are viewed on-disk, are long-observed but still not well-known structures in the solar atmosphere. Since they are outstanding features in multiple-wavelength observations of the Sun and have close relationships with various solar eruptive phenomena, prominences are always one of major topics in solar and space physics. The key issues in prominence/filament studies are their formation, maintenance, dynamic processes, and their roles in other related solar activities, e.g., coronal mass ejections (CMEs) and flares. With the aid of modern sensor technology, many facts of prominences are revealed \citep[e.g.,][and the references therein]{Poland_1986, Tandberg-Hanssen_1995, Martin_1998, Patsourakos_Vial_2002}. Prominences are dense (electron density $\sim10^{9}-10^{11}$ cm$^{-3}$) and cool ($\sim5000-8000$ K) plasmas  floating in hot and diluted solar corona \citep[e.g.,][]{Engvold_Brynildsen_1986, Hiei_etal_1986, Hirayama_1986, Madjarska_etal_1999}. They can appear anywhere from active regions to polar regions, and live for days to months. They have spines and barbs, and always straddle above polarity inversion lines. There are sometimes strong counterstreamings along spines and very dynamic vertical flows. The chirality of prominences/filaments obeys the pattern that most prominences on northern hemisphere are dextral while most ones on southern hemisphere are sinistral. The association rate of eruptive prominences with CMEs is more than about 70\% \citep[e.g.,][]{Gilbert_etal_2000, Gopalswamy_etal_2003}.

Many of the above findings are made through statistical investigations combined with case studies. A continuously updated catalog of prominences with unbiased parameters is undoubtedly helpful for such researches, especially in the age of the explosive growth of observational data. For instance, the successful launch of STEREO (Solar Terrestrial Relationship Observatory) spacecraft (A and B) in 2006 led the amount of solar observations explosively growing to more than 12 GB a day, and now it has increased to about 2 TB a day from SDO (Solar Dynamic Observatory), which was just launched in February 2010. NOAA/SWPC\footnote{http://www.swpc.noaa.gov/Data/index.html} routinely compiles a list of solar events, in which on-disk filaments and limb eruptive prominences are included; but the list is far from complete. Thanks to the unique properties of prominences/filaments, they can be clearly observed at multiple wavelengths, such as H$\alpha$, He I 10830~\AA, He II 304~\AA, radio waves, etc \citep[e.g.,][]{Schmahl_etal_1974, Hanaoka_etal_1994, Penn_etal_1994, ChiuderiDrago_etal_2001, Labrosse_Gouttebroze_2001}, and therefore  it is possible to  extract them from the vast amount of data automatically and consistently.

Recognitions of on-disk filaments and limb prominences are different. The former is mainly accomplished by studying H$\alpha$ data. For example, \citet{Gao_etal_2002}, \citet{Shih_Kowalski_2003}, \citet{Fuller_etal_2005} and \citet{Zharkova_etal_2005} developed codes to automatically detect filaments in full-disk H$\alpha$ images. The automated system developed by \citet{Bernasconi_etal_2005} is able to detect, classify and track H$\alpha$ filaments efficiently. EUV observations are much more difficult to be used in detecting on-disk filaments due to the low contrast of filaments and the involvement of coronal holes. However, EUV observations are suitable for limb prominence detection. Through the usage of Fe IX/X 171~\AA, Fe XII 195~\AA, Fe XV 284~\AA\ and  He II 304~\AA\ images from SOHO/EIT (\citet{Delaboudiniere_etal_1995}) instrument, \citet{Foullon_Verwichte_2006} developed algorithms to recognize limb prominences. In their method, He II 304~\AA\ data  provide the basic criteria for the selection of candidate prominence regions, and other emission lines are used to remove active regions, which also appear brightly in EUV 304~\AA. Most recently, \citet{Labrosse_etal_2010} also developed an efficient code to detect limb prominences in EUV 304~\AA\ images.

In this paper, we present an automated system of detecting and tracking solar limb prominences based on only He II 304~\AA\ data, and a resultant on-line catalog as well, which can be continuously updated. The performance and limitations of the system are presented in section \ref{sec_performance}. Based on our catalog, some preliminary statistical results of solar limb prominences are also presented. The reasons we choose He II 304~\AA\ emission line rather than H$\alpha$ are the following. First, for prominence/filament observations, He II 304~\AA\ line is the only one uninterruptedly imaging the Sun with high cadence (operated by space-borne instruments SECCHI/EUVI on board STEREO twins, and AIA on board SDO). A complete database of limb prominences is therefore possible to be established. Secondly, the high time resolution of the data allows us to track their evolution, even small changes. Thirdly, the projection effect can be minimized for certain  parameters, such as height, radial speed, etc. Fourth, they are complementary to the catalogs of on-disk filaments. Further, there is so far no well-established on-line catalog for limb prominences.

\section{Method}
Our system consists of five modules. The first module is to select prominence candidates; the second one is to extract necessary parameters for further usage; the third one is to discriminate prominences from other non-prominence features, such as active regions and noise; the fourth one is to track the prominences for the evolution; and the last one is to generate a catalog of prominences with final parameters. Here we use EUVI 304~\AA\ data from STEREO-B/SECCHI to illustrate these processes.

\subsection{Module 1: Prominence Candidate Selection}\label{sec_module_1}
\begin{figure*}[tbh]
  \centering
  \includegraphics[width=\hsize]{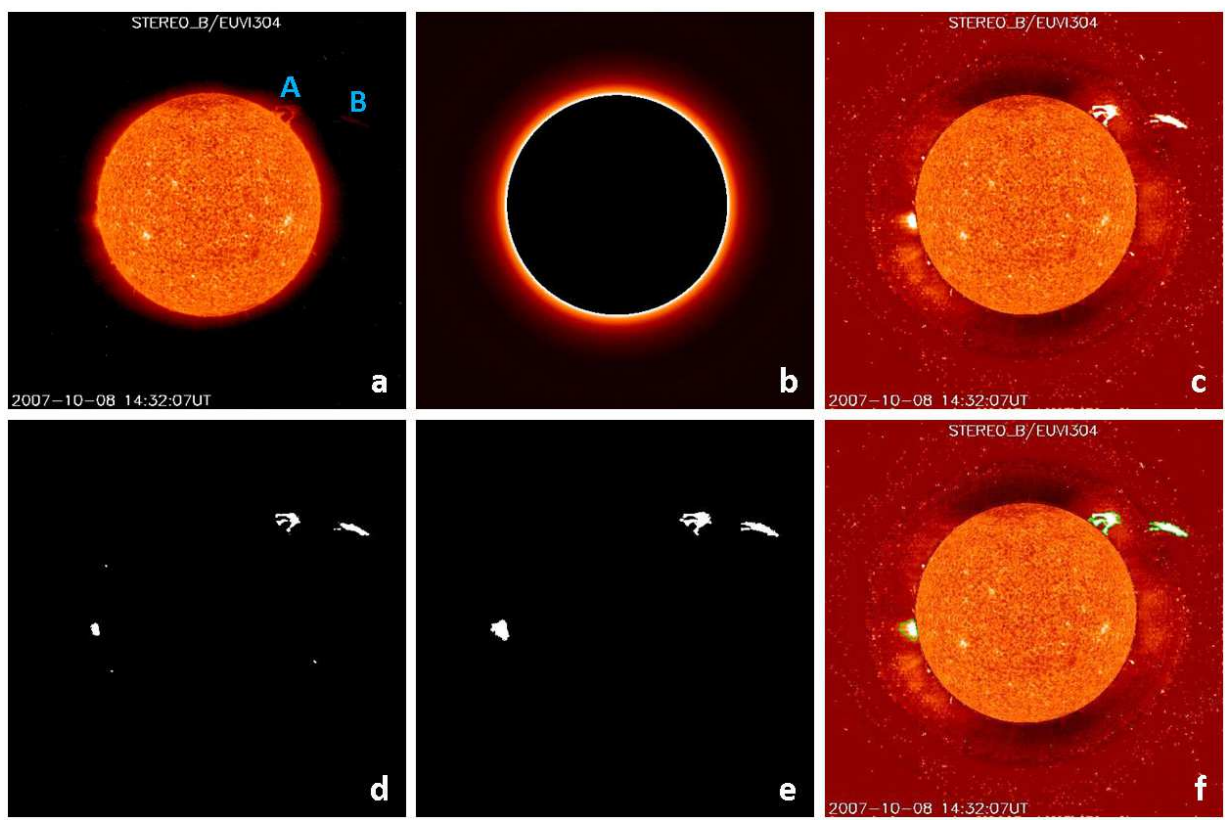}
  \caption{A sample image on 2007 October 8 to illustrate the processes of selecting prominence candidates.  (a) The raw  EUVI 304~\AA\ image,
(b) circular symmetrical background image, (c) rescaled image, (d) binary image of
selected kernels, (e) binary image of possible prominence regions,
and (f) the rescaled image with the boundaries of the recognized
regions. The bright patch above the east limb in (f) is an active
region, which will  be removed by module 3.}\label{fg_image_processing}
\end{figure*}

The functions of module 1 are illustrated in Figure \ref{fg_image_processing}. A raw EUVI 304~\AA image is shown in Figure \ref{fg_image_processing}a. The background brightness above the limb generally decreases with increasing distance, $r$, from the solar center. Similarly, the prominences near the solar surface are much brighter than those at high altitude (for example, comparing prominence A and B marked in the image). The variation of the brightness of the prominence with $r$ is further discussed in Sec.\ref{sec_fading}. Although region B is too dark to be noticed in the raw image, we still consider it a prominence candidate for its higher density compared to the ambient coronal plasmas.

The first step of processing is to use a technique similar to the normalizing-radial-graded filter \citep{Morgan_etal_2006} to rescale the  brightness so that the contrast is independent of $r$. To do this, a background image is first created, which is a circular symmetric image with respect to the center of the solar disk as shown in Figure \ref{fg_image_processing}b. The pixel value at any $r$ is just the average value of all the pixels along the circle at $r$ in the original image. It is obvious that the brightness of background plasma does drop quickly as $r$ increases. Then, we obtain  the rescaled image (Fig.\ref{fg_image_processing}c) by using the following formula
\begin{eqnarray}
\mathrm{Rescaled\ Image}=\frac{\mathrm{Original\ Image\ (Fig.\ref{fg_image_processing}a)}+\delta}{\mathrm{Background\ Image\ (Fig.\ref{fg_image_processing}b)}+\delta}
\end{eqnarray}
Here, $\delta$ is a small value to avoid dividing by a near-zero value. Both prominence material A and B become much clearer. For the STEREO-B/EUVI 304~\AA\ images, $\delta$ is  chosen to be 5 through trial and error; however, this number may change for other instruments, depending on the signal-to-noise ratio.

The further recognition, which applies the technique of region-growing with certain thresholds, is based on the rescaled image. The following process is similar to those by, e.g., \citet{Gao_etal_2002} and \citet{Bernasconi_etal_2005}, and thus we just briefly describe it here. First, we set a threshold $th_{knl}$ to pick all the pixels with larger value as kernels. The searching region is from 1 $R_\sun$ to $r_{max1}$, where $r_{max1}$ is an upper boundary, above which there is no kernel selected. The boundary of $r_{max1}$ is needed  because the signal-to noise ratio will become low when approaching the edge of the telescope's field of view. For STEREO-B/SECCHI EUVI images, we choose the value of 1.7 $R_\sun$. The selected kernels serve as the seeds, from which the whole prominence regions grow out. Figure \ref{fg_image_processing}d is a binary image showing the  kernels. Here some small kernels, which are isolated pixels due to the presence of noise, have been removed by applying a morphological {\it opening} operator with a box size of $s_n\times s_n$. Secondly, let these kernels grow by setting another smaller threshold $th_{pro}$, i.e., all neighboring pixels with values larger than $th_{pro}$ will be included in the growing  regions. For the cases that several regions close to each other but not connected, we use a morphological {\it closing} operator with a box size of $s_m\times s_m$ to merge them together. Some regions, whose areas are smaller than $th_{area}$, are discarded to further prevent noise-like features from being included. The resultant regions are the candidate prominences as shown in Figure \ref{fg_image_processing}e. Figure \ref{fg_image_processing}f is obtained by superimposing the boundaries of the recognized regions on the rescaled image. The set of arguments discussed above are listed in Table \ref{tb_arguments}.

\subsection{Module 2: Parameter Extraction}\label{sec_module_2}
Once we have the boundary of a region of interest, the extraction of parameters of the region is straightforward. According to the scaling information in the header of FITS  file of the image, we can calculate  the area ($A$) and average brightness ($F$) of the region, the minimum and maximum positions in both radial and azimuthal directions ($r_{bot}$, $r_{top}$, $\theta_{min}$, and $\theta_{max}$), and the centroid of brightness ($r_{cen}$ and $\theta_{cen}$). It should be noted that, as the signal-to-noise ratio decreases significantly near the edge of the field of view, we set another upper boundary $r_{max2}$, slightly larger than $r_{max1}$, and consider that the parameters of any prominence extending into the region above it might be unreliable.

\begin{figure}[tbhp]
  \centering
  \includegraphics[width=0.5\hsize]{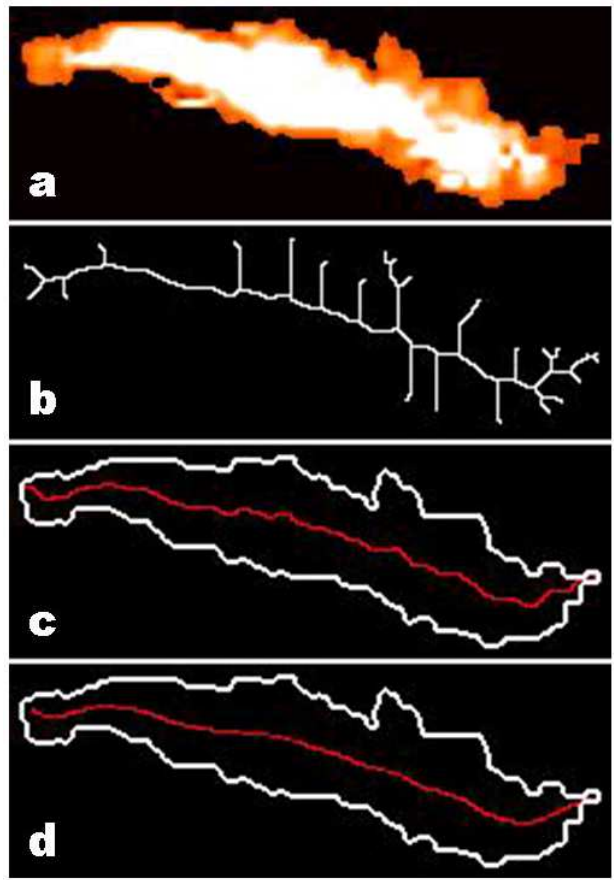}
  \caption{Extracting the spine of a prominence. (a) Original region, (b) Skeleton, (c) Spine, (d) Smoothed spine.}\label{fg_skeleton}
\end{figure}

Further, we linearize (i.e., get the spine of) the region to obtain certain morphological information. Figure \ref{fg_skeleton} presents a sample. We use the morphological {\it thin} operator \citep[refer to, e.g.][]{Lam_etal_1992} to get the skeleton (panel b) of the prominence of interest (panel a). Usually, a skeleton is too intricate because of many branches involved. To remove trivial branches, we first calculate the length (weighted by the rescaled brightness) of each branch. Then, for branches connecting to the same node, we compare their lengths and save the largest one. The above steps are iterated until there are only two ends (panel c). The resultant curve is further smoothed to get the spine (panel d) by applying the 2-dimensional mean filter method. One should note that any region of interest will be finally simplified to a line with only two ends even if it actually has three or more ends/footpoints.

Since most prominences are loop-like structure in morphology, we keep the length of the spine as the characteristic length ($L$) of the recognized region. Meanwhile, the obtained spine can be used in 3D reconstruction of a prominence if it is viewed in two different visual angles at the same time (e.g., by the STEREO twins or combined with SOHO), which will be specifically studied in another paper.

\subsection{Module 3: Non-prominence Feature Removal}
While photons in EUV 304~\AA\ images mainly come from He II emissions, they also have contaminations from hot coronal lines \citep[e.g.,][]{Zhang_etal_1999}. As a result, prominences are not the only bright feature in EUV 304~\AA\ wavelength,  and active regions also appear bright. In \citet{Foullon_Verwichte_2006} work, the authors realized this fact and used observations in other wavelengths to exclude active regions from their detected bright regions. The regions recognized through our first two modules also contain active regions and  some  noisy features. We do not, however, try to involve other observations in our detection, which will make the system more intricate and prone to additional errors. In our system, the previously extracted parameters for each recognized region will be used to discriminate real prominences from these non-prominence features, as discussed below.

Prominences have a different appearance from other features. For example, in morphology, a prominence usually looks like a loop or stick, while an active region is shaped as a round blob. In brightness, a prominence is almost flat over radial distance in the rescaled image, while an active region is not. Thus one can use a classification method to remove the non-prominence features. There are many classification methods, e.g., linear discriminant analysis (LDA), support vector machines (SVM), neural networks (NN) \citep[e.g.,][]{Meyer_etal_2003}. The method we adopted here is the linear discriminant analysis. One can refer to the paper by, e.g., \citet{Fisher_1936} for the principle of the linear discriminant analysis.

Through many tests, the parameters $\ln A$ (standing for the size of a region), $\ln\frac{A}{L}$ (for the shape) and $\ln \chi_F^2$ (for the variation in brightness, where $\chi_F^2$ is the value of Chi-square goodness-of-fit statistic for the brightness $F$ as a linear function of distance $r$) are chosen to construct the linear discriminant function (LDF). Our sample contains 5066 regions from a total of 3780 images (4 images per day, near 00:00, 06:00, 12:00, 18:00 UT, respectively, from 2007 April 1 to 2009 October 31). Each region is checked by eyes to determine which group it belongs to, the prominences or non-prominences. On the basis of this large collection of features of known classification, or the truth table, we derive the LDF as
\begin{eqnarray}
X=1.460\ln A+1.103\ln\frac{A}{L}-0.491\ln \chi_F^2 \label{eq_lda}
\end{eqnarray}
or
\begin{eqnarray}
X=14.20\frac{\ln A}{\langle\ln A\rangle}+4.70\frac{\ln\frac{A}{L}}{\langle\ln\frac{A}{L}\rangle}+1.36\frac{\ln \chi_F^2}{\langle\ln \chi_F^2\rangle}
\end{eqnarray}
where the quantity $\langle f\rangle$, the mean value of $f$ calculated based on our truth table, is used to normalize the parameters, and so that we can learn the importance of the parameters from their factors. The area $A$ is the most important to discriminate a prominence from other features because it has the largest factor 14.20. However, it should be noted that some very big prominences might be missed due to the important role of area in the discrimination (for example the erupting prominence on 2009 November 2). But such missings usually take place in several frames, and therefore will not significantly affect the tracking result of the whole evolution process of the prominences identified in other frames.

\begin{figure}[tbh]
  \centering
  \includegraphics[width=0.9\hsize]{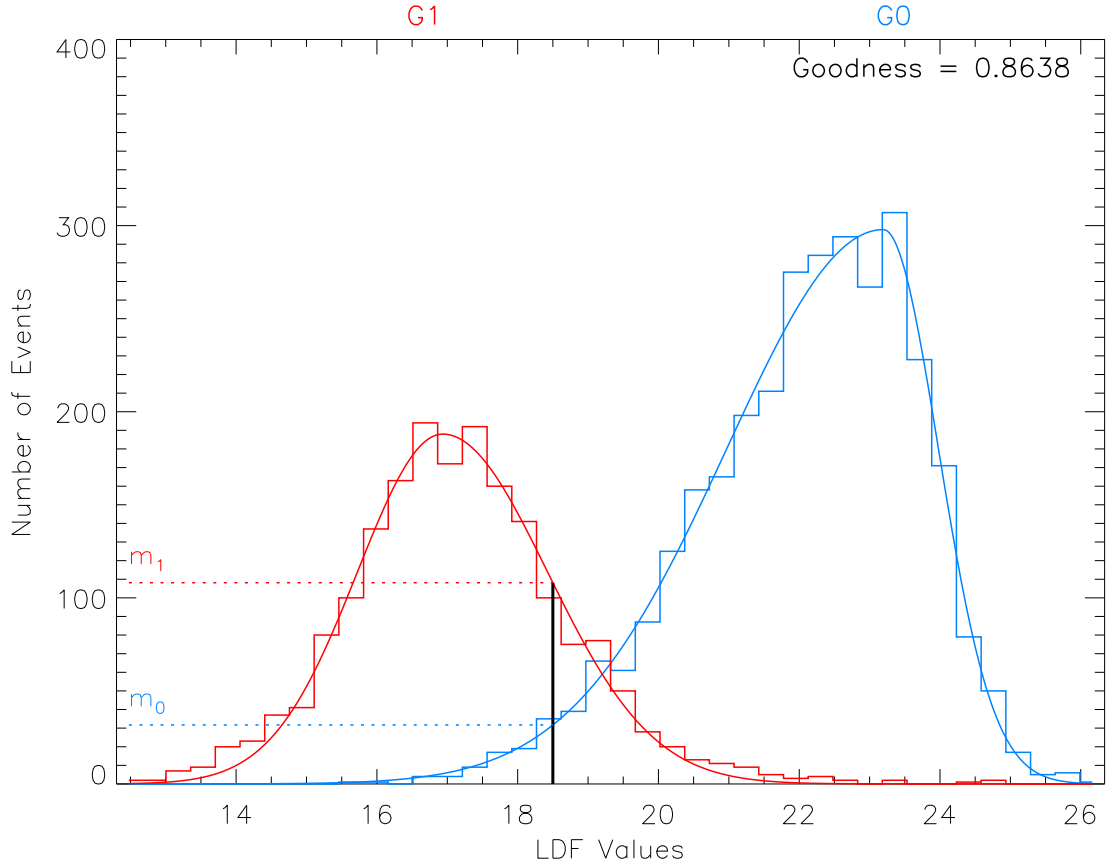}
  \caption{Result of the linear discriminant analysis of the truth table. The two groups, prominences (labeled G1) and non-prominence
features (G0), are indicated in red and blue colors,
respectively.}\label{fg_lda}
\end{figure}

Figure \ref{fg_lda} shows the discriminant result. It can be seen that the group of prominences (labeled G1) generally have different LDF values from the group of non-prominence features (G0). Since there is still an overlap between the two groups, we evaluate the goodness of LDF by
\begin{eqnarray}
G=1-\frac{n_o}{n}
\end{eqnarray}
where $n_o$ is the number of regions whose LDF value falls within  the overlap, and $n$ is the total number of regions in the truth table. In other words, the value of $G$ is the ratio of the area of non-overlapped regions to the sum of the areas occupied by the two groups in Figure \ref{fg_lda}. $G=1$ means the LDF can completely discriminate the two groups. In our case, the goodness is about 0.86.

Based on Eq.\ref{eq_lda}, we can calculate the LDF  value of any recognized region, and compare it with the derived distribution of the LDF values, which is fitted with Gaussian distribution functions as shown by the curves in Figure \ref{fg_lda}, to determine how likely the region is a prominence. The likelihood of a region being a prominence is given by
\begin{eqnarray}
P=\frac{m_1}{m_0+m_1}
\end{eqnarray}
where $m_0$ and $m_1$ are the values of the fitted Gaussian distribution functions corresponding to the LDF value for group 0 and 1, respectively. A region with $P\leq50\%$ is treated as a non-prominence feature and discarded.

\subsection{Module 4: Prominence Tracking}\label{sec_module_4}
\begin{figure*}[tbh]
  \centering
  \includegraphics[width=\hsize]{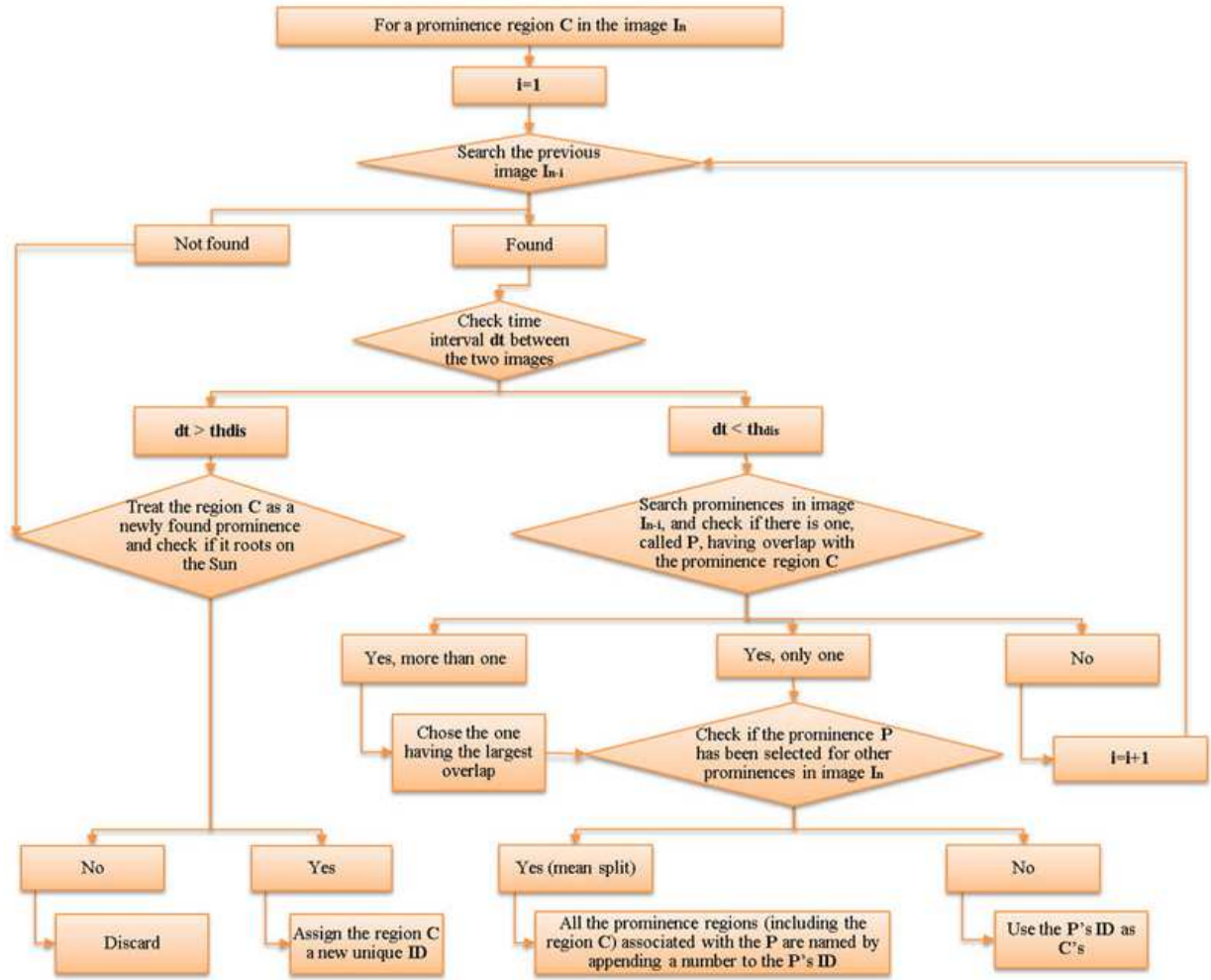}
  \caption{Flow chart to illustrate the prominence tracking process.}\label{fg_tracking}
\end{figure*}

Our method to track the evolution of a prominence is quite simple. Figure \ref{fg_tracking} is the flow chart showing how we track a prominence. The top of the flow chart is a prominence to be tracked in an image, and the bottom of the chart gives the four possible results. Since the flow chart is detailed enough, we will not repeat it here. There are only several things that we would like to point out.

\begin{figure*}[tbh]
  \centering
  \includegraphics[width=0.32\hsize]{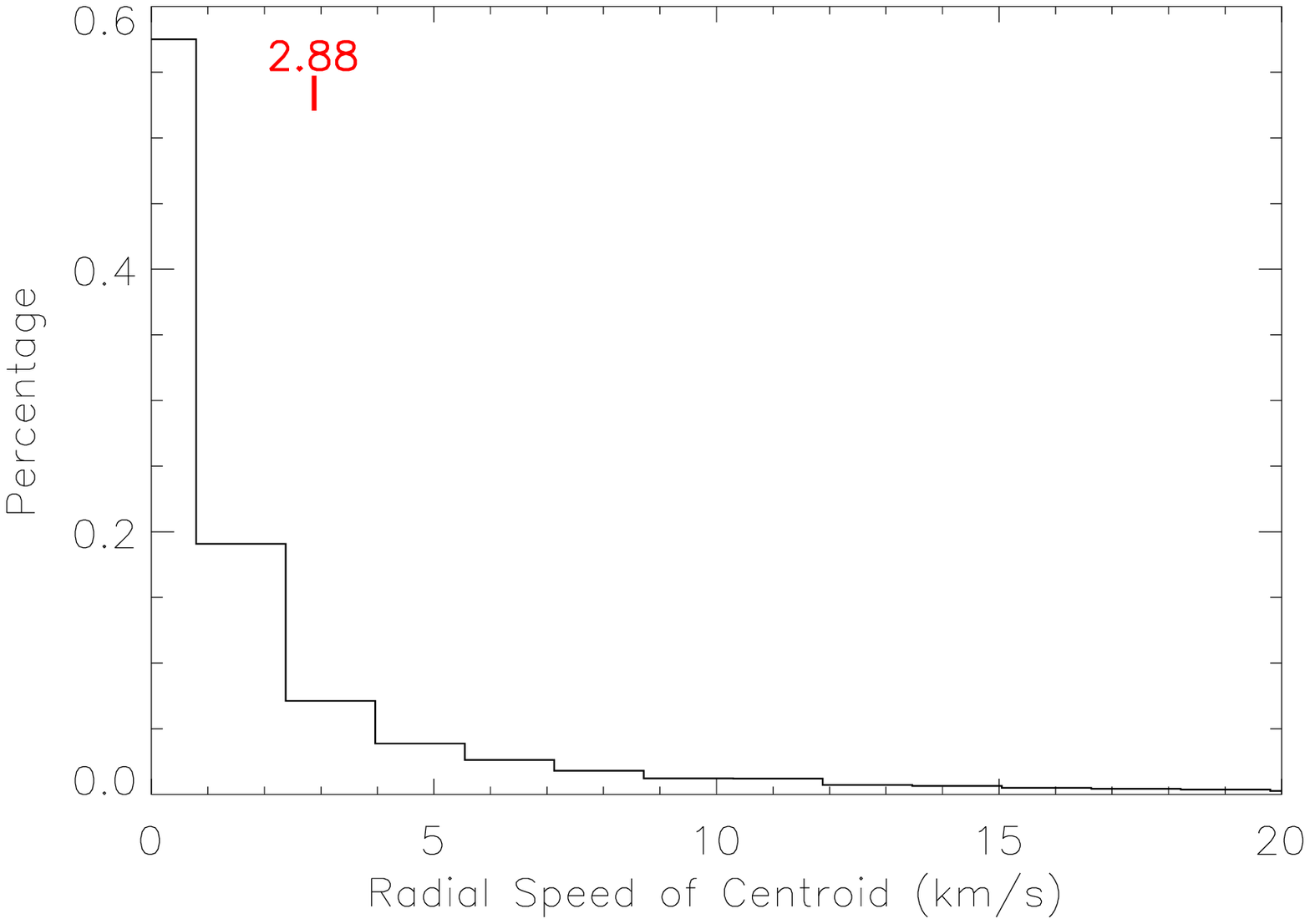}
  \includegraphics[width=0.32\hsize]{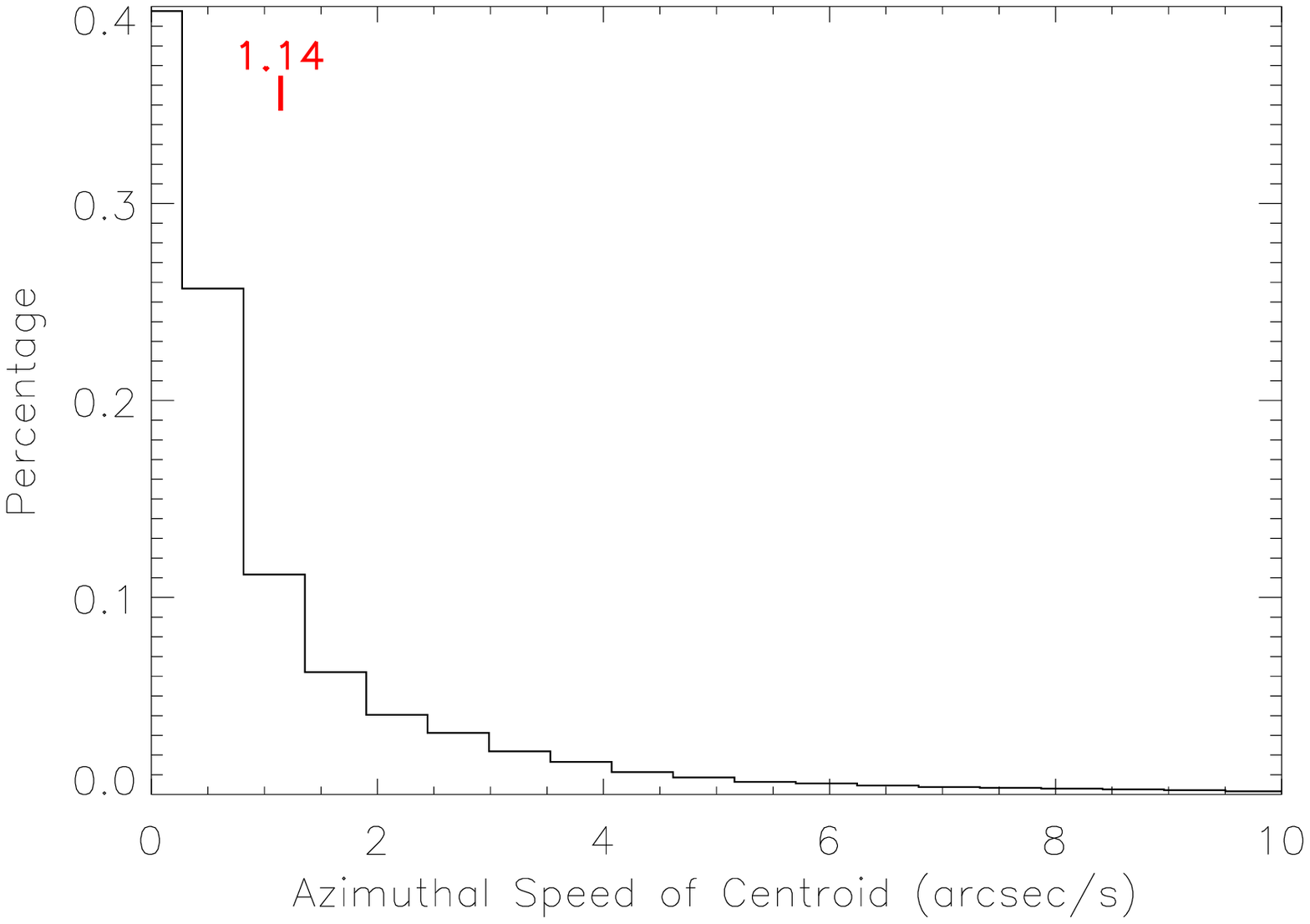}
  \includegraphics[width=0.32\hsize]{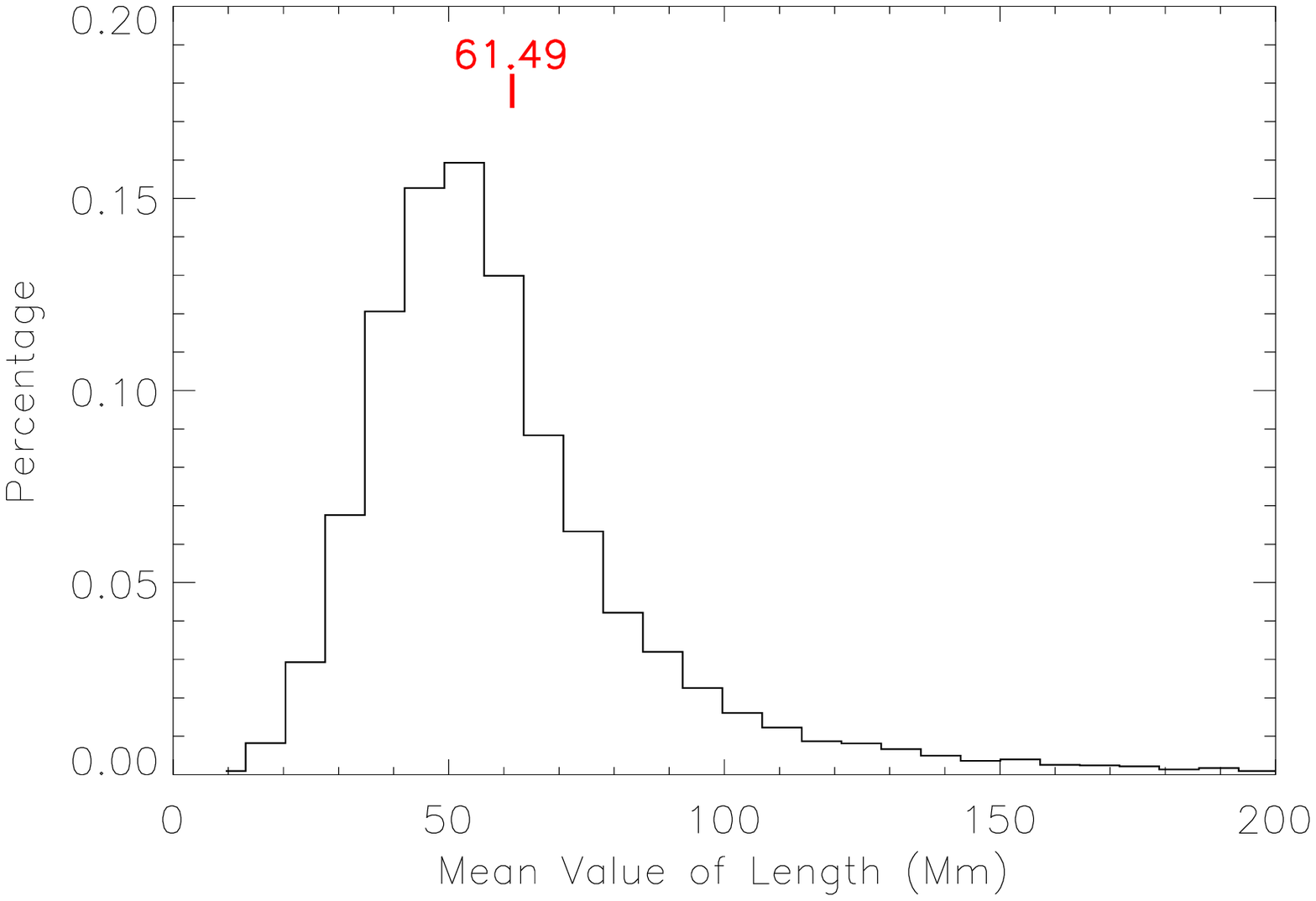}
  \caption{Histograms of the radial ({\it left panel}) and azimuthal speed ({\it middle panel}) of the centroid of prominences and the characteristic length ({\it right panel}). The average values are marked in the plots. The upper limits of the $x$-axes are chosen to make the plots readable, but do not mean the maximum values (The same treatment is made to the Fig.\ref{fg_duration} and \ref{fg_compare}).}\label{fg_cadence}
\end{figure*}

First, the criterion used to judge if a prominence region is evolved from a prominence region in the previous image is to check whether or not there is an overlap between them in spatial domain. This requires that the cadence of the data should be high enough, especially when studying a fast erupting prominence. According to our statistical result (see Fig.\ref{fg_cadence}), which is plotted based on our catalog (refer to Sec.\ref{sec_catalog}), most prominences move with a speed of about 4 km/s or less in radial or azimuthal direction, few ones may reach up to more than one hundred km/s. Considering that their characteristic length is $L\approx 60$ Mm, it is inferred that a cadence of 4 hours (for 4 km/s speed, or a cadence of 15 minutes for 100 km/s) is basically sufficient for prominence tracking, which is much longer than 10 minutes, the cadence of STEREO/SECCHI EUVI 304~\AA\ data.

Secondly, we use a time threshold $th_{dis}$ to determine whether or not a prominence has disappeared, i.e., if a previous named prominence has not been found in the successive images for a duration of $th_{dis}$, it is treated as disappeared. A previous detected prominence may temporarily and intermittently `disappear'. Such a `disappearing' may not be real; it may be resulted from the unstable quality or jittering of images (although we have done some treatments on original data as mentioned in Sec.\ref{sec_module_1}) and/or small changes of prominence itself, which cause the brightness of the prominence to decrease down below the threshold $th_{knl}$ or even $th_{pro}$ temporarily. Note that this situation only happens to some small and/or faint prominences, not to major ones. Set a relatively long time duration $th_{dis}$ can efficiently track the entire evolution process of a prominence. Here we let $th_{dis}=2$ hours.

Thirdly, in our tracking process, the case that a prominence splits into two or more parts is considered (see the third result in the flow chart). However, we do not deal with the case of merging, which means there are more than one prominence regions (for example, A and B) in the previous image merging together and associated with only one prominence region (say C) in the current image. The merging of prominences is ambiguous, as the  phenomenon can also be interpreted as that the prominence region A (or B) disappears and region B (or A) evolves to region C. In this scenario, no region merging takes place.

Fourth, if a prominence is identified as a new one (see the left side of the flow chart), we will check if it connects to the solar surface. Only those rooted on the Sun are considered as real prominences. This justification is based on the assumption that no newly emerged prominence is disconnected from the Sun.

\subsection{Module 5: Catalog Generating}\label{sec_catalog}
\begin{table*}[t]
\caption{List of the arguments used by SLIPCAT for STEREO-B/EUVI
304~\AA\ data} \label{tb_arguments}
\begin{tabular}{lccp{285pt}}
\hline
Arguments & Values &Units & Interpretation  \\
\hline
$\delta$ $^a$   & 5.0       & & A small value used in creating rescaled images (see Sec.\ref{sec_module_1}). \\
$th_{knl}$ $^a$ & 2.0     & & A threshold for kernel selection (see Sec.\ref{sec_module_1}). \\
$th_{pro}$ $^a$ & 1.7  & & A threshold for region growing (see Sec.\ref{sec_module_1}). \\
$r_{max1}$ $^b$ & 1.7 & $R_\sun$ & An upper boundary in $r$, above which there is no selected kernel (see Sec.\ref{sec_module_1}). \\
$r_{max2}$ $^b$ & 1.73 & $R_\sun$ & An upper boundary in $r$. The parameters of any prominence extending into the region above it is considered to be unreliable (see Sec.\ref{sec_module_2}). \\
$s_n$ $^c$ & 5 & pixels & Define a box used to remove noise-like kernels (see Sec.\ref{sec_module_1}). \\
$s_m$ $^c$ & 5 & pixels & Define a box used to merge regions which are very close to each other (see Sec.\ref{sec_module_1}). \\
$th_{area}$ & 500 & Mm$^2$ & A threshold to remove very small regions (see Sec.\ref{sec_module_1}). \\
$th_{dis}$ & 2 & hours & A threshold to judge if a prominence has disappeared (see Sec.\ref{sec_module_4}). \\
\hline
\end{tabular}\\
$^a$ Pixel values in rescaled images, probably changing for different instruments.\\
$^b$ Depending on the range of field of view and the signal-to-noise ratio. \\
$^c$ Need to be changed for different spatial resolution of images.
\end{table*}

\begin{table*}[t]
\caption{List of the primary parameters extracted for each prominence at a certain time}
\label{tb_parameters}
\begin{tabular}{lp{395pt}}
\hline
Parameter & Interpretation  \\
\hline
($r_{cen}$, $\theta_{cen}$) & Coordinates of the centroid of the brightness.  \\
$r_{bot}$, $r_{top}$ & Give the span of a prominence in the radial direction.   \\
$\theta_{min}$, $\theta_{max}$ & Give the span of a prominence in the angular direction.  \\
$A$ & Area of a prominence in the units of Mm$^2$.  \\
$L$ & Characteristic length of a prominence in the units of Mm.  \\
$F$ & Average brightness recorded in the original image.  \\
$P$ & Likelihood of a recognized region to be a prominence. \\
\hline
\end{tabular}
\end{table*}

Solar prominences above the limb are identified  and tracked by the above four modules. All the input arguments to SLIPCAT for the STEREO-B/SECCHI EUVI 304~\AA\ data are summarized in Table \ref{tb_arguments}, and the primary parameters extracted for each prominence at a certain time are summarized in Table \ref{tb_parameters}. Since we have the parameters of each prominence in its time sequence, it becomes possible to automatically extract  its kinetic evolution information. For example, we can derive the velocity $\ve v_c$ and acceleration $\ve a_c$ of the centroid, from the variations of $r_{cen}$ and $\theta_{cen}$ with time. Also we can get some peak and average values of the above parameters, such as $A_{max}$, $A_{ave}$, etc. Moreover, for each prominence, we give the confidence level by the following formula
\begin{eqnarray}
C=\left\{\begin{array}{lc}
1, & \overline{P}>90\% \\
2, & 75\%<\overline{P}\le 90\% \\
3, & 50\%<\overline{P}\le75\% \\
\end{array}\right.
\end{eqnarray}
where $\overline{P}$ is the average value of the likelihood of the prominence over its period of tracking. A resultant online catalog is established at \url{http://space.ustc.edu.cn/dreams/slipcat/}, where one can find all the final output parameters. The  analyses in the following sections   are based on the parameters in the catalog.

We note that it is straightforward  to apply the system to other spacecraft data, e.g., SDO/AIA data (which is in our planning), though we present here only STEREO-B data. The only part we need to modify is the input arguments listed in Table \ref{tb_arguments}. Besides, the code is written in IDL language, and the modules 1-3 and modules 4-5 can run separately. The modules 1-3 do nothing with the causal relationship among the images, thus they can process images in serial or parallel.  Using a computer with 2.33 GHz Inter Xeon CPUs, 3 GB memory and Linux operating system, it takes 36 seconds on average for modules 1-3 to process one full-size EUVI image and takes only 0.29 seconds for modules 4-5. Thus, to process the sequence of images over one day with the cadence of 10 minutes , it needs about 1.45 hours. The SDO/AIA data has higher resolution ($4096\times4096$ pixels) and faster cadence (10 seconds), which will  require modules 1-3 to spend much more time. It  is estimated that it will take about 347 hours if we serially process such one-day images in the  machine mentioned above. Thus, to  reduce the processing time to less than 24 hours,  there is a need of a small cluster with 15 or more CPUs to run the modules 1-3 in parallel, which should be affordable.

\section{Performance and Limitations}\label{sec_performance}
\begin{figure*}[tbhp]
  \centering
  \includegraphics[width=0.49\hsize]{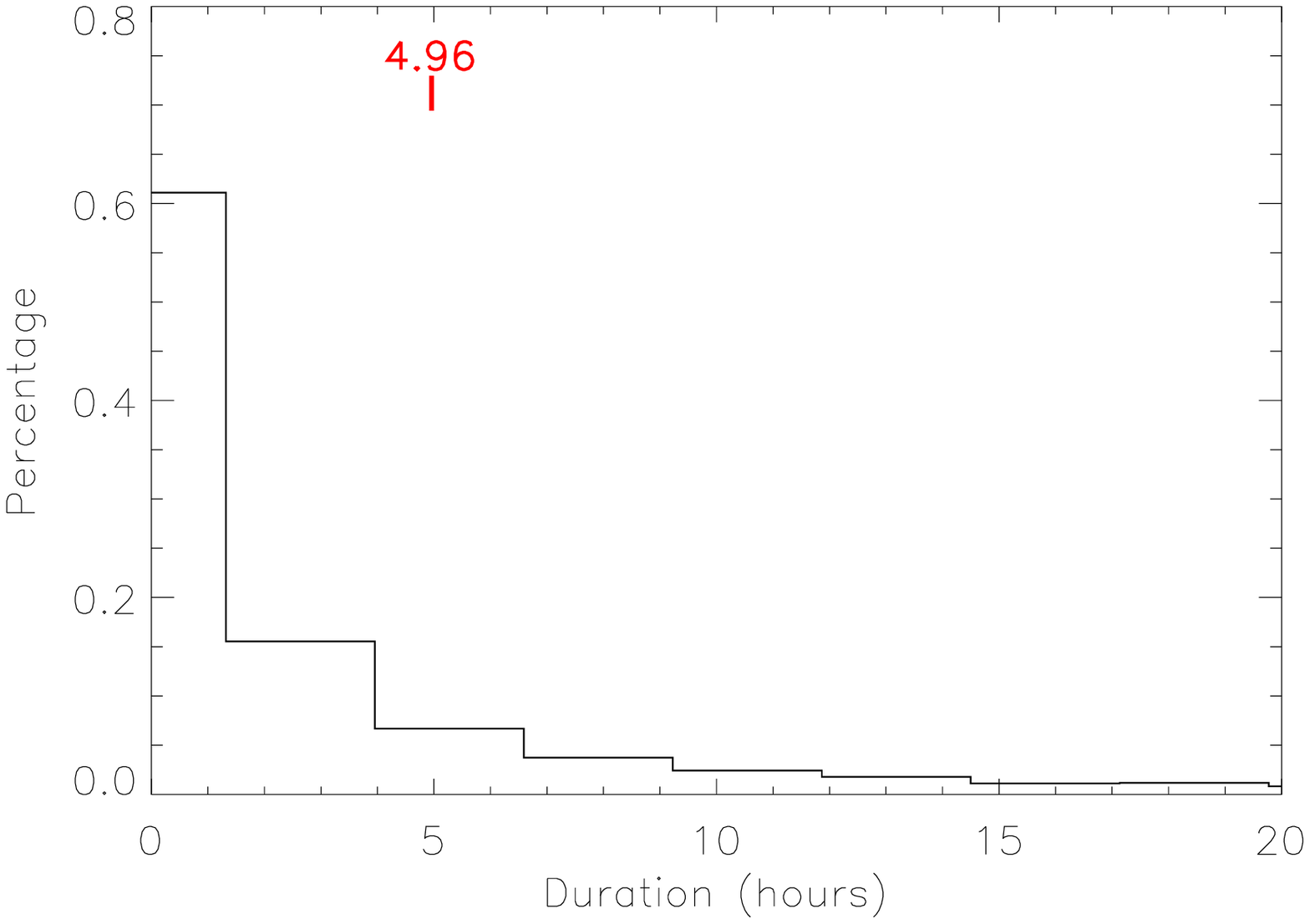}
  \includegraphics[width=0.49\hsize]{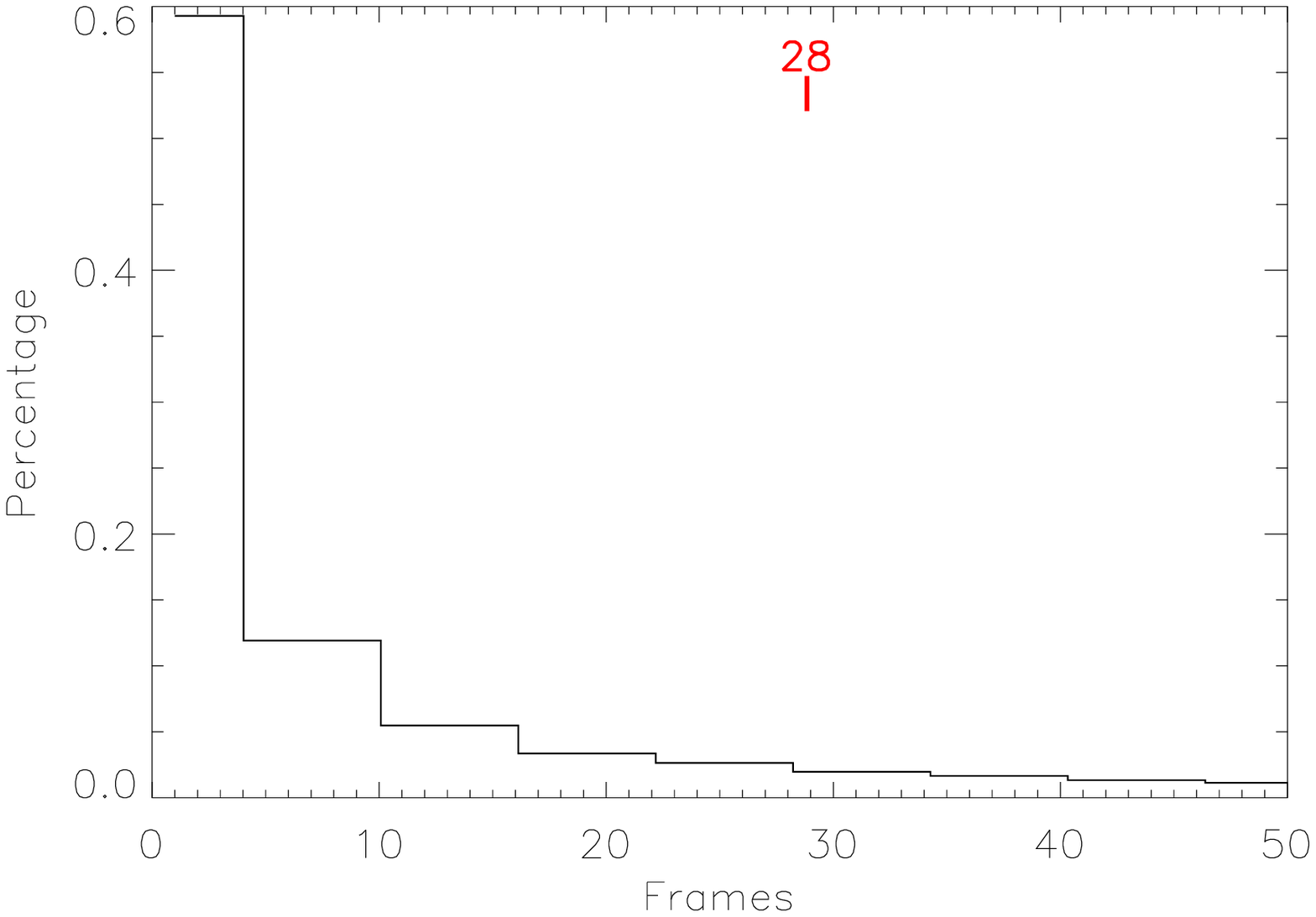}
  \includegraphics[width=0.98\hsize]{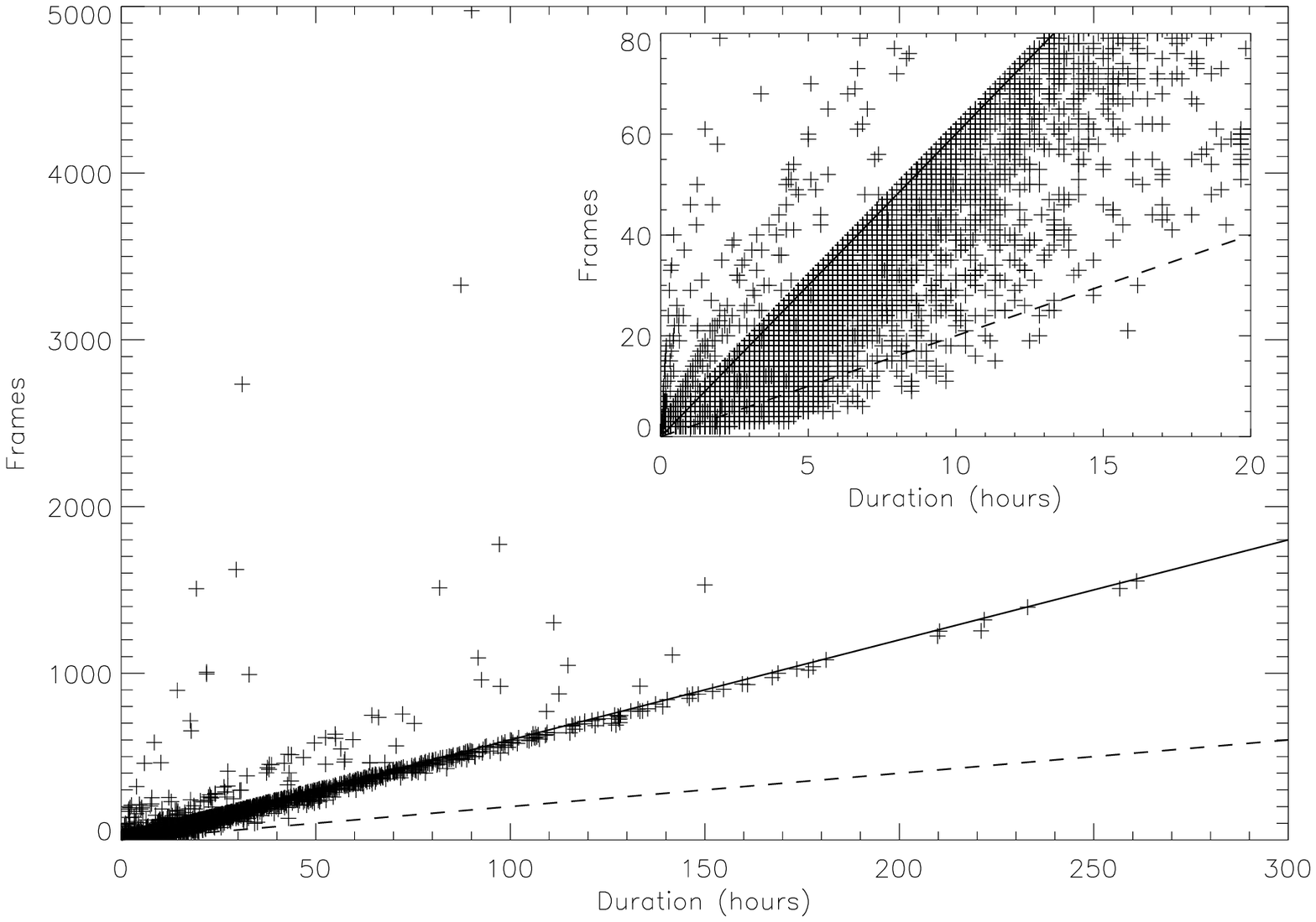}
  \caption{({\it Upper panels}) Histograms of the duration and the number of frames. The average values are marked in the plots.
({\it Lower panel}) Relationship between the duration and frames.
The solid line indicates the expected relationship at a cadence of
10 minutes.}\label{fg_duration}
\end{figure*}

The SLIPCAT detects 19140 prominences from the STEREO-B data during the beginning of April 2007 to the end of October 2009. Figure \ref{fg_duration} shows the distributions of duration of the prominences and the number of frames in which a prominence is detected. It is found that 6348 prominences (33\%) are detected in only one frame (upper panels), and the rest exhibits a roughly linear correlation between the duration and frame number as shown in the lower panel of Fig.\ref{fg_duration}. The solid line in the scatter-plot marks the expected relationship between the duration and frame number for the imaging cadence of 10 minutes. A point above the line means the instrument operates in a higher-cadence mode and the prominence is well tracked, while a point below the line implies that the prominence is missing in some frames. Most points distribute around the solid line. We define a prominence as poorly-tracked when it matches one of the following two criteria. (1) It is detected in only one frame, and (2) it is missing in 2/3 or more of the expected frames (marked by the dashed line in the scatter-plot). Note that the radial speeds of prominences are no more than 160 km/s (see Fig.\ref{fg_cadence}), which implies a prominence is expected in at least 5 or more frames. Thus a prominence detected in only 2 frames is also treated as poorly-tracked. There are a total of 9663 (50.5\%) poorly-tracked prominences during the period of interest. As to the rest, we call them well-tracked prominences.

\begin{figure*}[tbh]
  \centering
  \includegraphics[width=0.32\hsize]{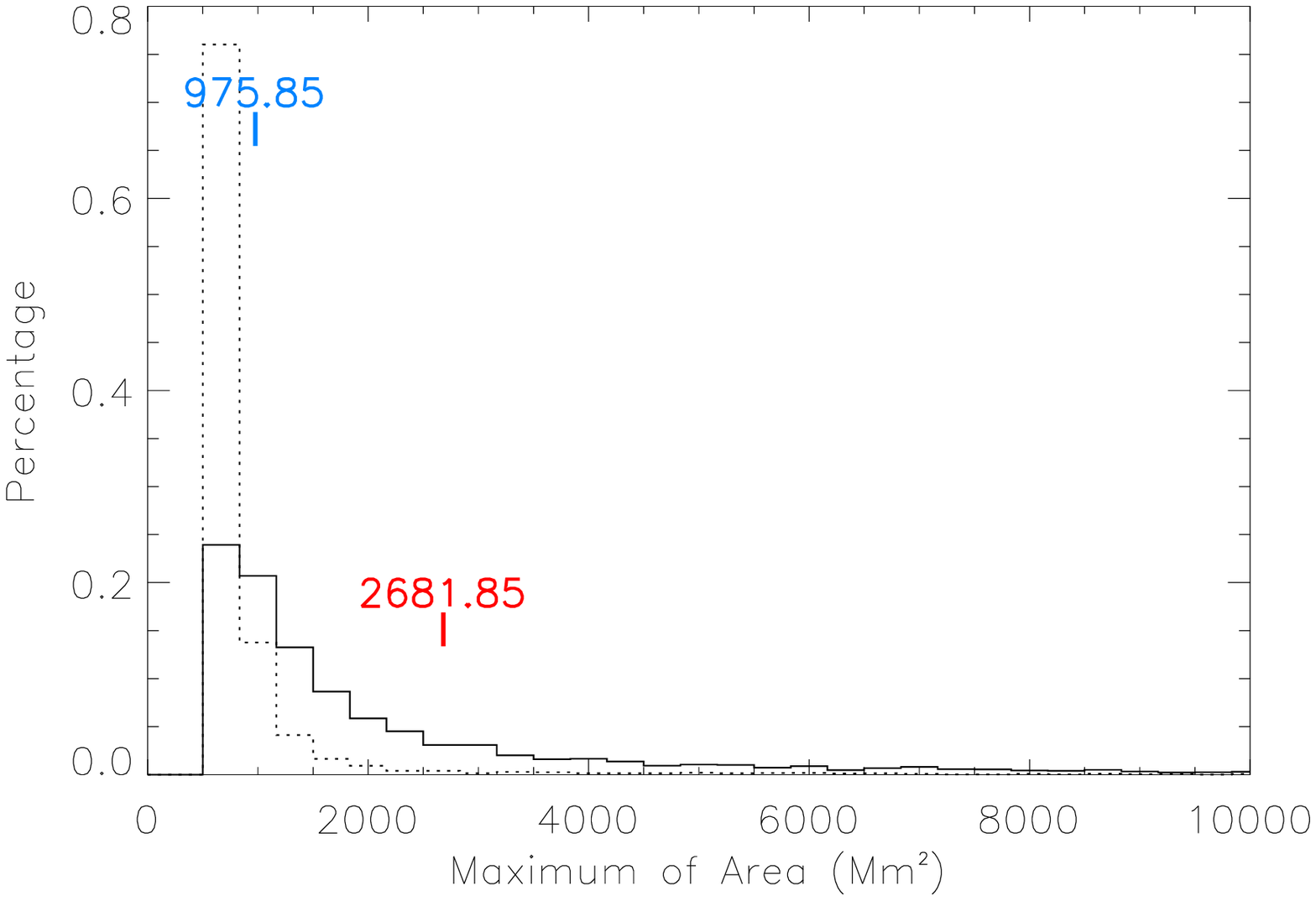}
  \includegraphics[width=0.32\hsize]{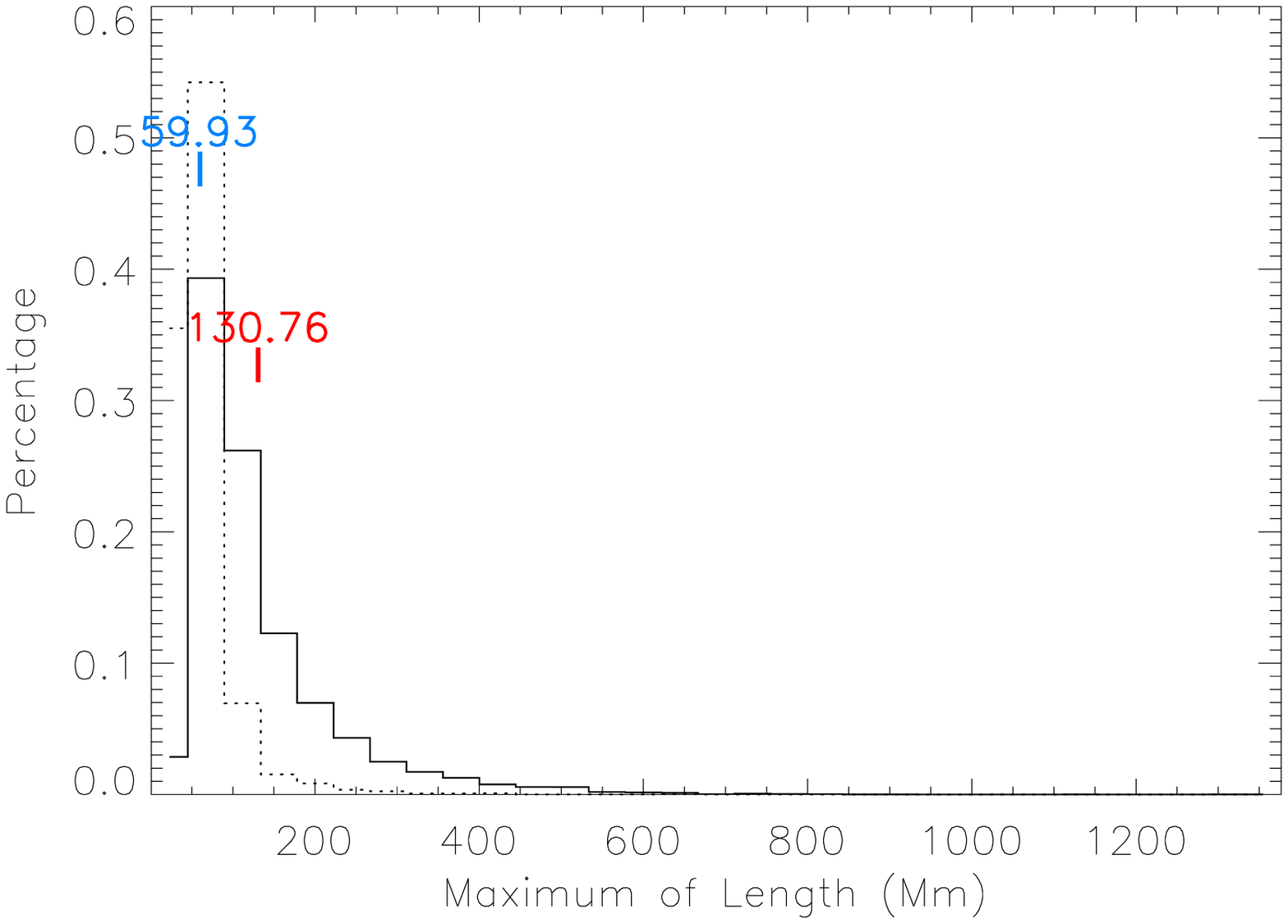}
  \includegraphics[width=0.32\hsize]{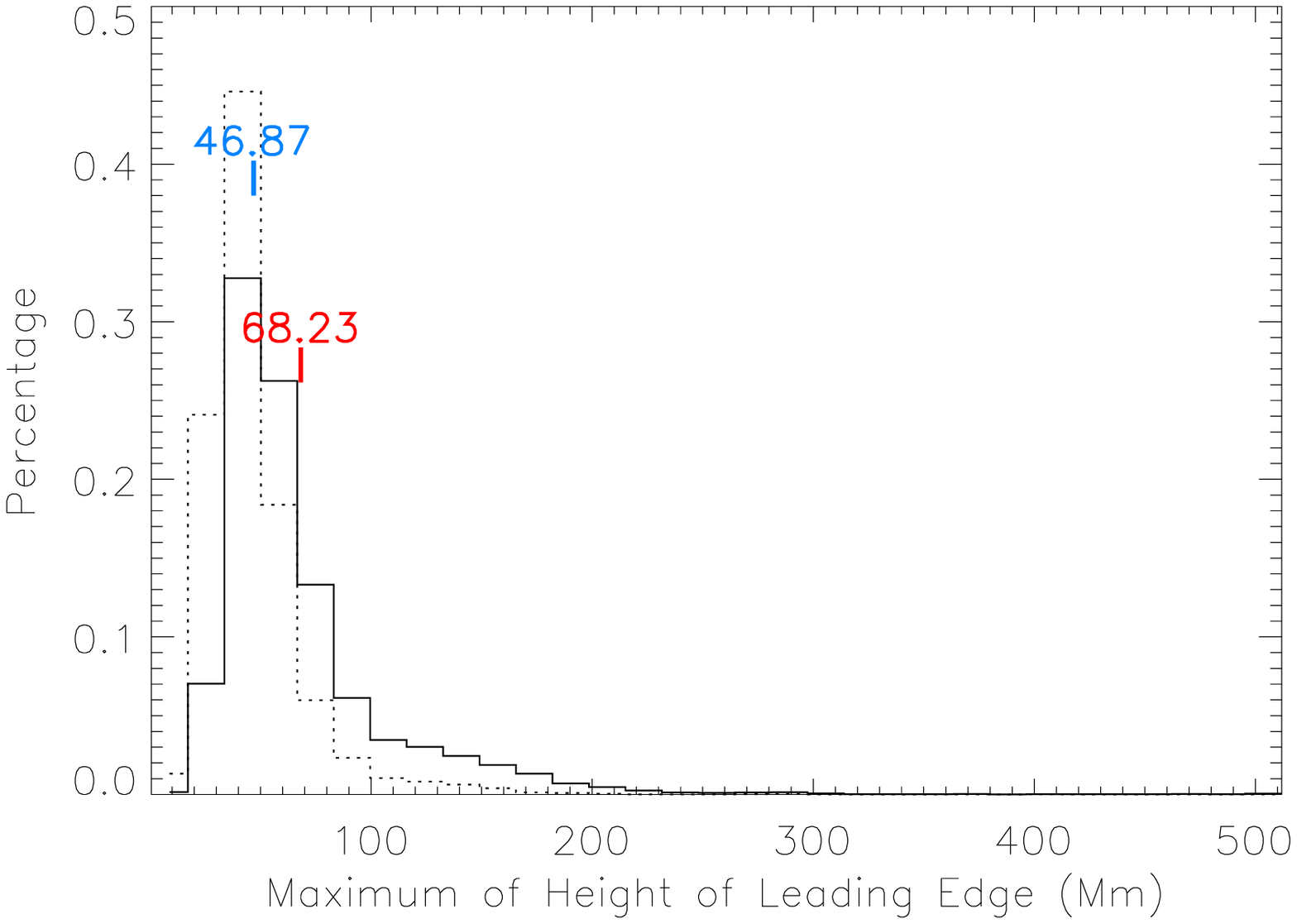}
  \caption{Histograms of the maximum area ({\it left panel}), characteristic length ({\it middle panel}) and top position ({\it right panel}) for the well-tracked (solid line) and poorly-tracked (dotted line) prominences. The average values (red for well-tracked and blue for poorly-tracked) are marked in the plots.}\label{fg_compare}
\end{figure*}

These poorly-tracked prominences are generally small and their top portions (or leading edges) lie low. These are revealed in Figure \ref{fg_compare}. The average value of maximum areas of poorly-tracked prominences is 976 Mm$^2$, about 3 times smaller than that of well-tracked ones; the average value of maximum lengths of poorly-tracked prominences is about 60 Mm, nearly half of that of well-tracked ones; and the average top position of poorly-tracked prominences is about 21 Mm lower than that of well-tracked ones. As early as in 1932, \citet{Pettit_1932} had concluded that prominences are usually about 60 Mm long or more, 10 Mm thick and 50 Mm high. These numbers are close to what we obtained here  using modern data. By checking movies in the catalog, we find that the causes of such poorly-tracked prominences are probably resulted from the features to be marginal (close to the detection thresholds),  contamination of nearby non-prominence features, and of course, the limitation of the detection algorithm. The parameters of these prominences may not be extracted correctly, and therefore they will be excluded from our statistical analysis in Sec.\ref{sec_statistics}.

\begin{figure}[ptbh]
  \centering
  \includegraphics[width=0.8\hsize]{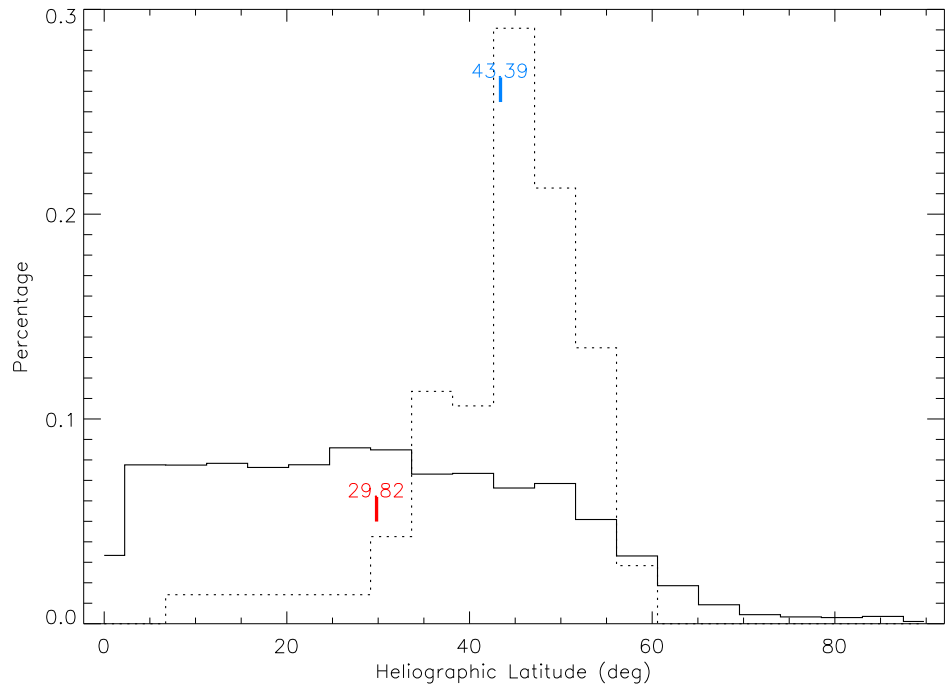}
  \caption{Histograms of the latitude for long ($\geq80$ hours, dotted line) and short ($<80$ hours, solid line) duration prominences, respectively. The average values (red for short-duration and blue for long-duration) are marked in the plots.}\label{fg_latitude}
\end{figure}

\begin{figure}[ptbh]
  \centering
  \includegraphics[width=0.8\hsize]{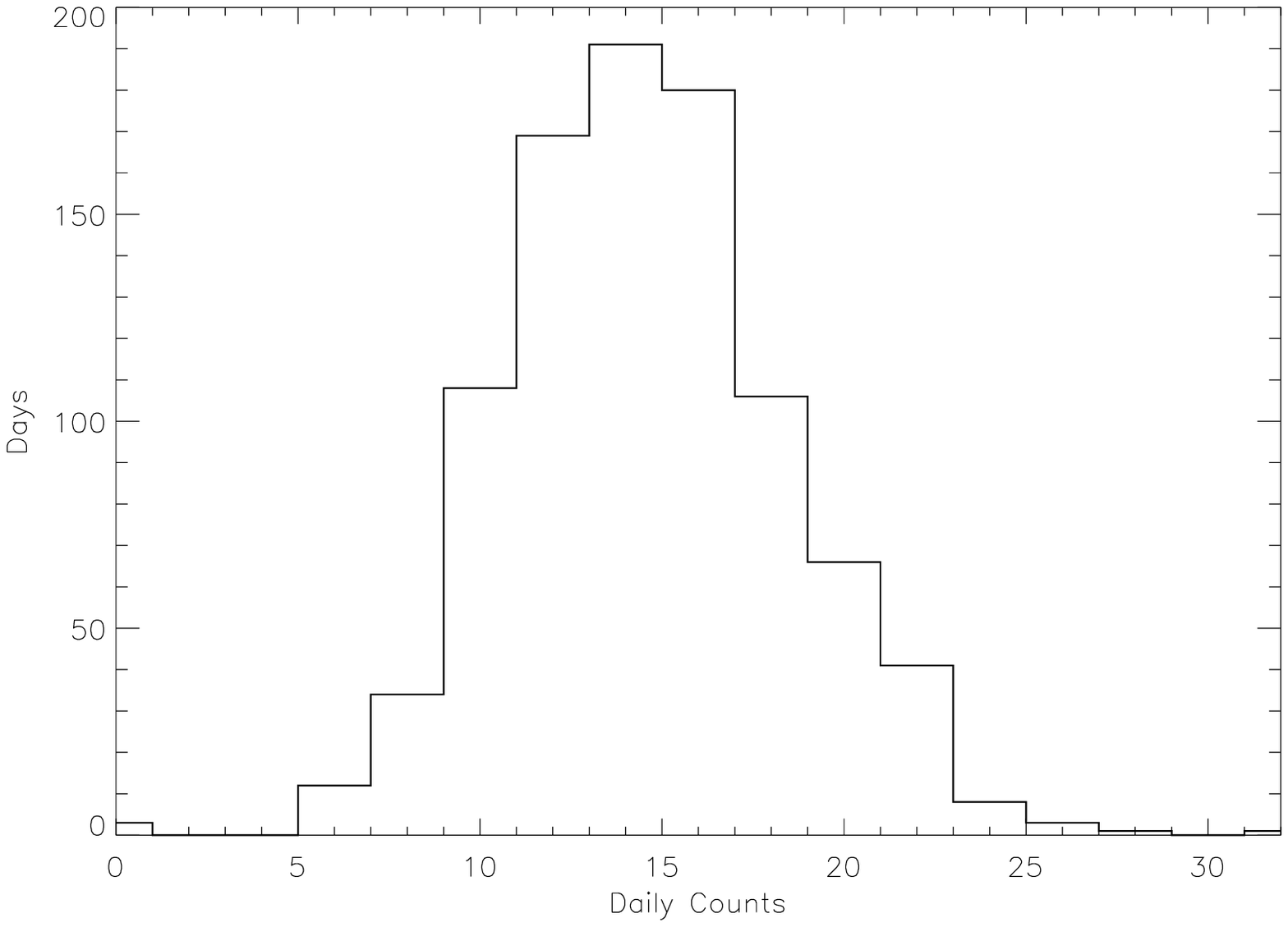}
  \caption{Distribution of daily counts of limb prominences.}\label{fg_daily_counts}
\end{figure}

The upper-left panel of Figure \ref{fg_duration} suggests that some prominences may appear for up to 260 hours, which seems too long to be possible. Due to the solar rotation, a prominence at a height of about 56 Mm (the average value of top position) above equator can stay visible for no more than 80 hours in theory. With the increasing latitude, this lasting duration may increase. If the prominence extends along the longitudinal  direction, the lasting duration can be even longer. Figure \ref{fg_latitude} shows that the prominences with long duration generally appear in high latitude, where quiescent or polar crown prominences are usually present. Thus, it is possible to have such long-duration prominences.

\begin{figure*}[tbh]
  \centering
  \includegraphics[width=\hsize]{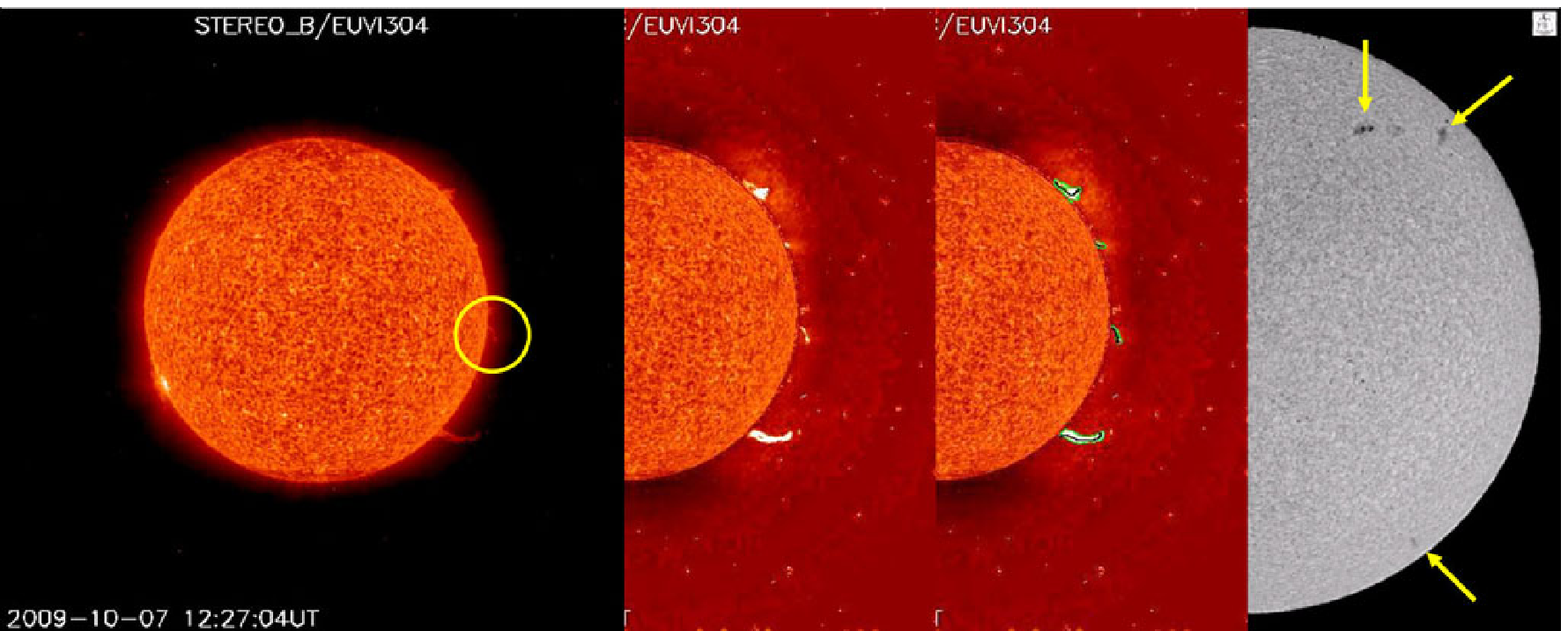}
  \caption{Comparison of the EUV 304~\AA\ prominences viewed by STEREO with H$\alpha$ filaments viewed from the Earth.
From the left to right, they are original EUV 304~\AA\ image,
rescaled image, rescaled image with selected prominences, and
H$\alpha$ image from BBSO.}\label{fg_ha_slipcat}
\end{figure*}

Figure \ref{fg_daily_counts} shows the daily counts of the well-tracked prominences. For most days, about 14 prominences are detected. However, in extreme cases, there may be as many as 32 or as few as zero prominences a day. By checking the H$\alpha$ images from BBSO, it is found that there are more filaments during solar maximum than solar minimum. During 2007 -- 2009, the extreme solar minimum, there are usually only a few filaments in an H$\alpha$ image, that is inconsistent with our results. To make sure that the prominences identified by SLIPCAT are not non-prominence features, we compare the EUVI 304~\AA\ images with the H$\alpha$ image as shown in Figure \ref{fg_ha_slipcat}. The date of these images is 2009 October 7th, which is arbitrarily chosen; on this day, the STEREO-B spacecraft was 58 degrees behind the Earth. Thus the west limb in STEREO-B/EUVI 304~\AA\ image corresponds to about 33 degrees west to the central meridian in the H$\alpha$ image. One can find that the western hemisphere in the H$\alpha$ image is largely free of features except there are three prominences standing at high latitude (as denoted by arrows). Actually, there is a prominence at the low latitude, which can be clearly seen in the EUVI images (marked by a circle). The comparison demonstrates that there are probably some small prominences visible in EUV 304~\AA\ emission line, but invisible in H$\alpha$. It is also supported by many observational facts that solar filaments are generally more extended in EUV lines than in H$\alpha$ \citep[e.g.,][]{Heinzel_etal_2001}. Thus, it can be concluded that SLIPCAT is sensitive to recognize prominences, even those invisible in H$\alpha$ .

\begin{figure*}[ptbh]
  \centering
  \includegraphics[width=\hsize]{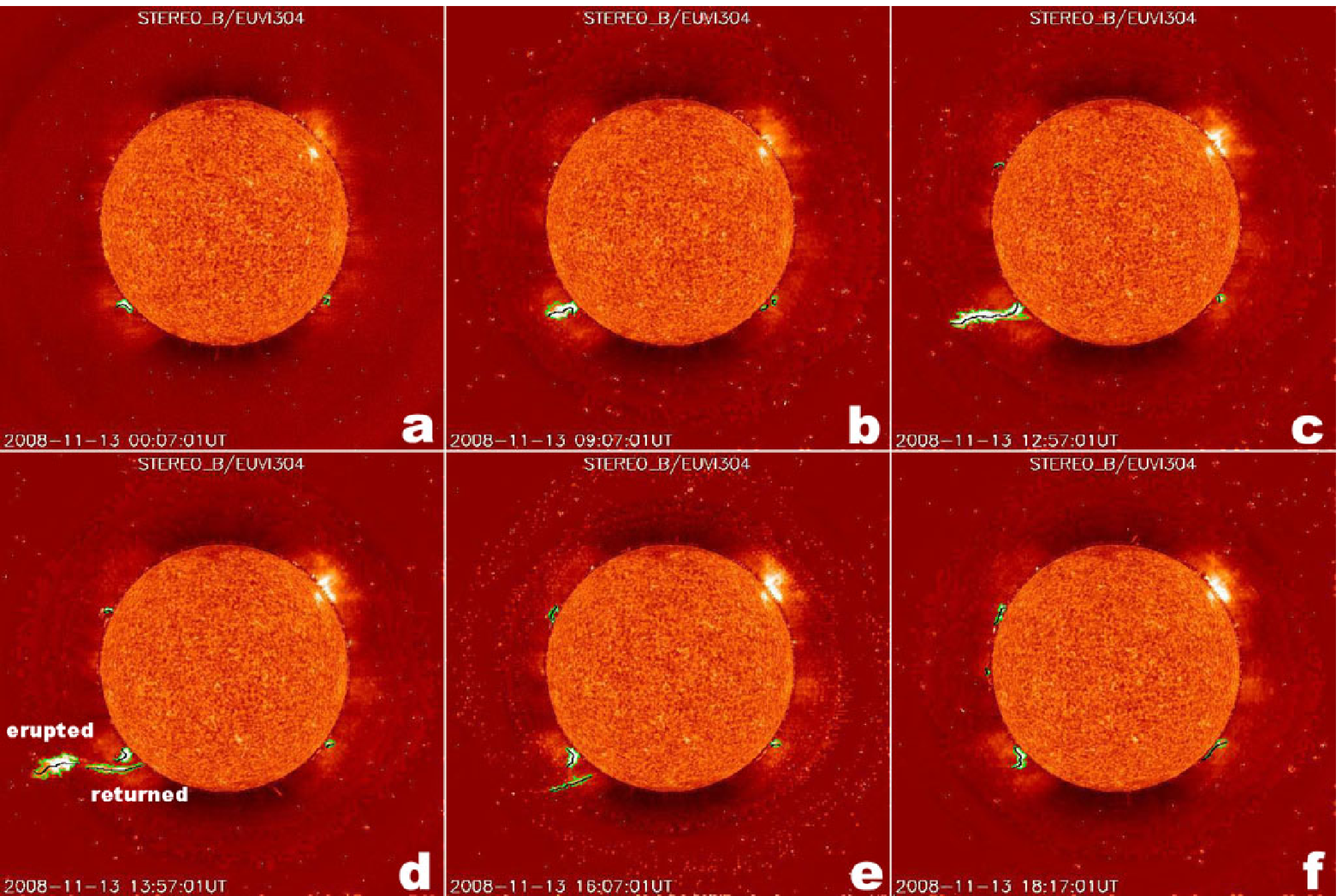}
  \caption{An erupting prominence on 2008 November 13, which split
  into three parts during the eruption. One escaped from the Sun, one
  remained on the Sun, and the other erupted but returned back
  to the Sun.}\label{fg_20081113_prom}
\end{figure*}

\begin{figure*}[ptbh]
  \centering
  \includegraphics[width=0.495\hsize]{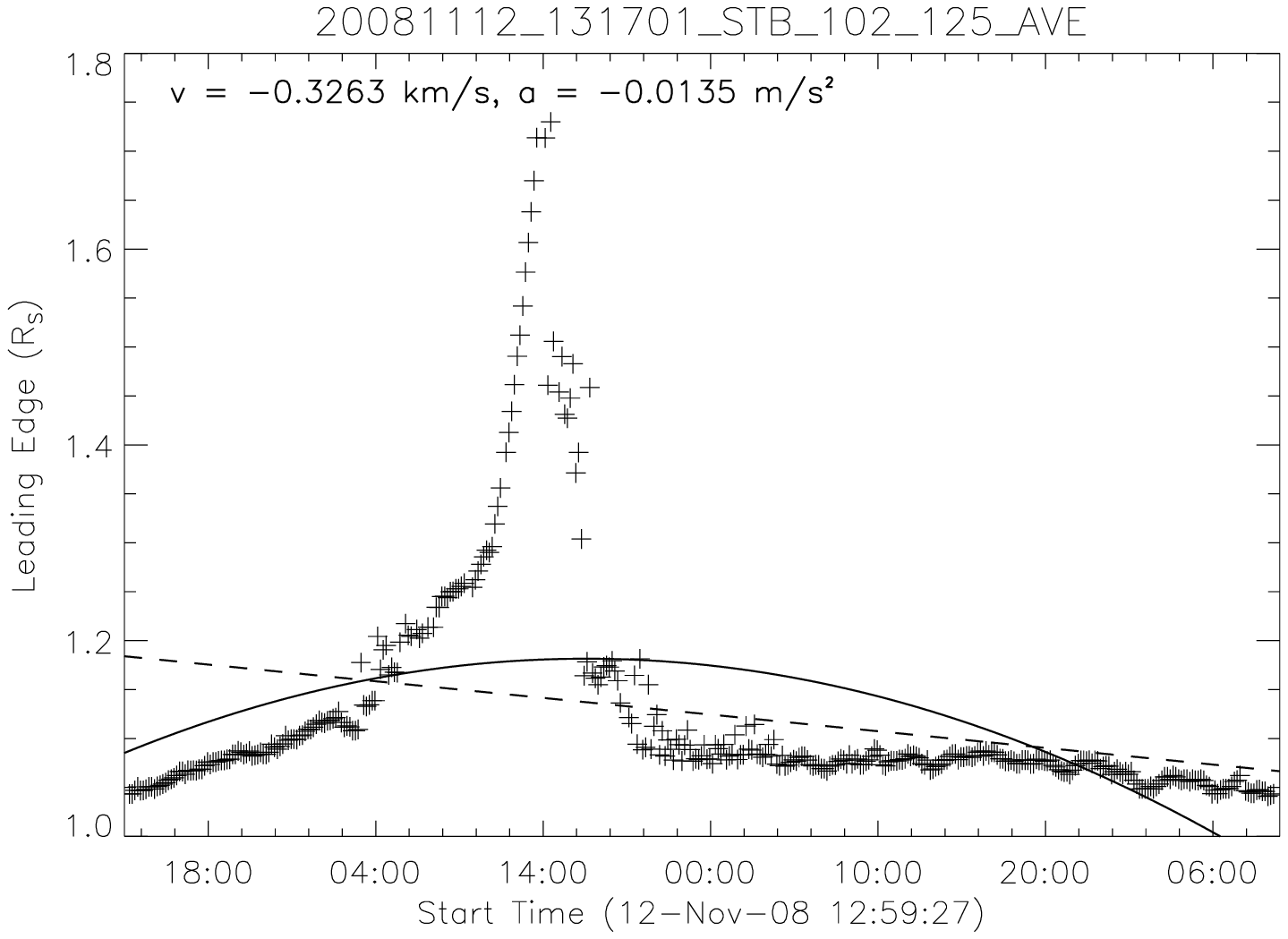}
  \includegraphics[width=0.495\hsize]{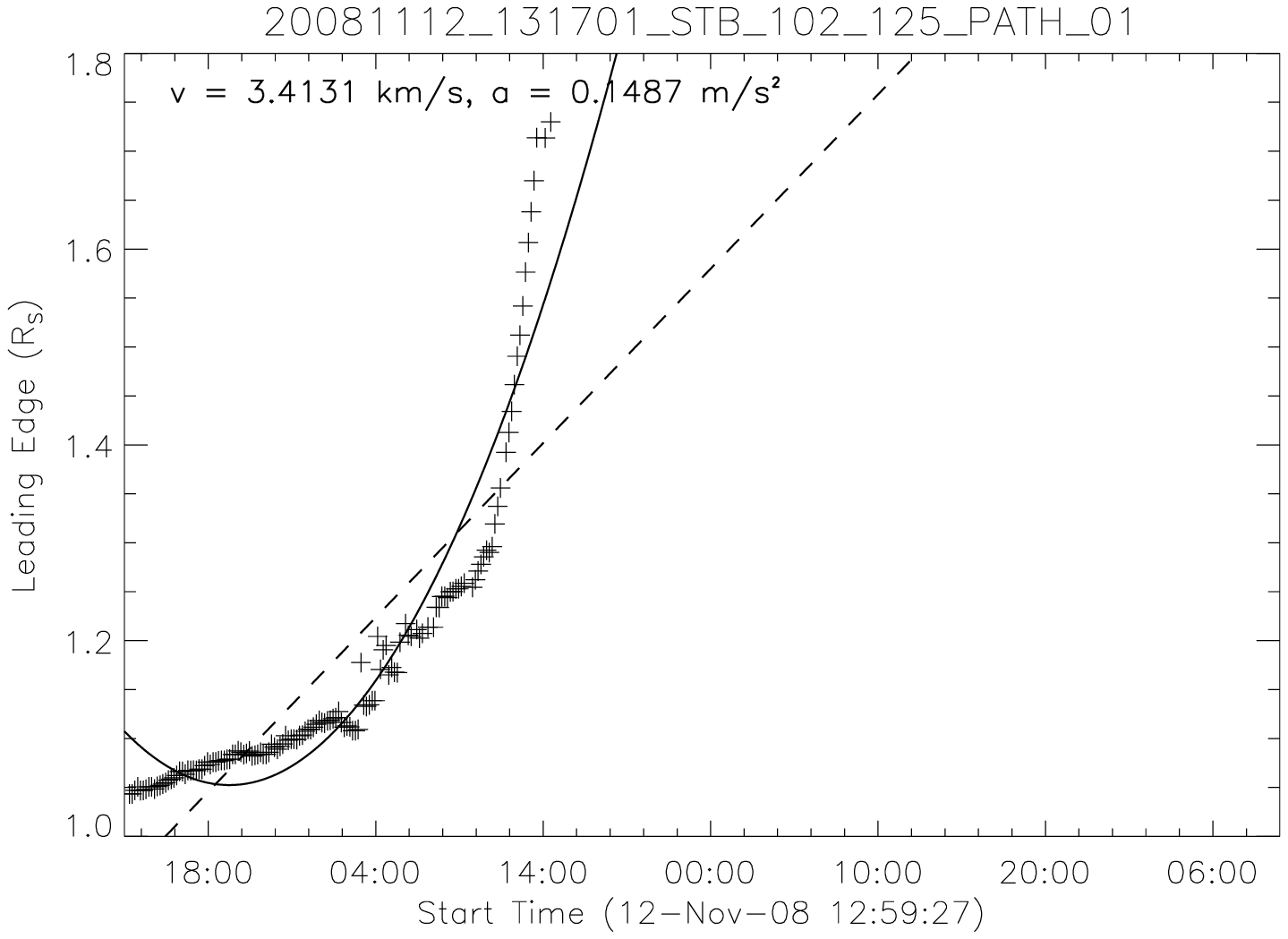}
  \caption{Height-time profiles of the leading edge of the prominence presented in Fig.\ref{fg_20081113_prom}.
  The left panel is for the whole prominence system, while the right panel is for the escaping part only. The solid line is a quadratic fit through the data points and the
dashed line is a linear fit.}\label{fg_20081113_kin}
\end{figure*}

Further, the accuracy of the parameters listed in the catalog should be addressed. First of all, it should be noted that the values of these parameters just give us the information of prominences during the period they are detected, not necessarily through their whole life-time. Second, some parameters suffer from the projection effect, e.g., the value of area depending on the angle of view. Third, some parameters, e.g., velocity, change rates of area and brightness, are automatically derived from fitting data points. Thus their accuracy depends on the number of frames, the fitting function and the complexity of the prominence. As an example, Figure \ref{fg_20081113_prom} presents an erupting prominence observed on 2008 November 13. This prominence rose from the south-east limb, and partially erupted. During its eruption, the prominence split into 3 major parts (see Fig.\ref{fg_20081113_prom}d), one escaped from the Sun, one remained on the Sun, and the other erupted but returned back to the Sun later. SLIPCAT tracks the whole eruption process. The left panel of Figure \ref{fg_20081113_kin} displays the height-time profile of the leading edge of the prominence. The solid line is a quadratic fit through the data points and the dashed line is a linear fit. Due to the splitting of the prominence, the fitting curves obviously do not reflect the reality. The correct treatment is to study the evolution of the split parts separately. The right panel shows the height-time profile of the escaping part. However, the fitting result for that part is still not satisfactory enough. The time at about 11:00 UT is a critical point, before which the escaping part was slowly rising, while after which it experienced a fast acceleration and erupted quickly. Thus a two-stage linear fitting is more appropriate than a pure linear or quadratic fitting. However, we still choose  linear and quadratic functions to fit all the detected prominences, because this treatment can be easily operated in an automated way. One should keep it in mind that the fitting results only give a coarse estimation of the average speed (or change rates of area and brightness) and acceleration.

\section{Preliminary Statistical Results}\label{sec_statistics}
Now SLIPCAT has a complete dataset for STEREO-B data. A catalog of prominences seen from STEREO-A will be generated soon. Here we would like to present some statistical results of the STEREO-B prominences. Since we probably do not get very accurate parameters for prominences as discussed in the last section, the results obtained here are just preliminary.  However, from a statistical point of view, these results should be significant. In the following analysis, all the poorly-tracked prominences are removed. Such prominences are generally extremely small and stay at low altitude as discussed in Sec.\ref{sec_performance}, thus the statistical results might suffer from  the bias of selection; one  should treat it as a statistics of moderate and  major prominences. Moreover, we will include prominences with all confidence levels. There are only 257 (2.7\%) prominences with confidence level of 2 or 3, which represents a rather small fraction in the database .

\subsection{Static Parameters}
\begin{figure*}[ptbh]
  \centering
  \includegraphics[width=0.49\hsize]{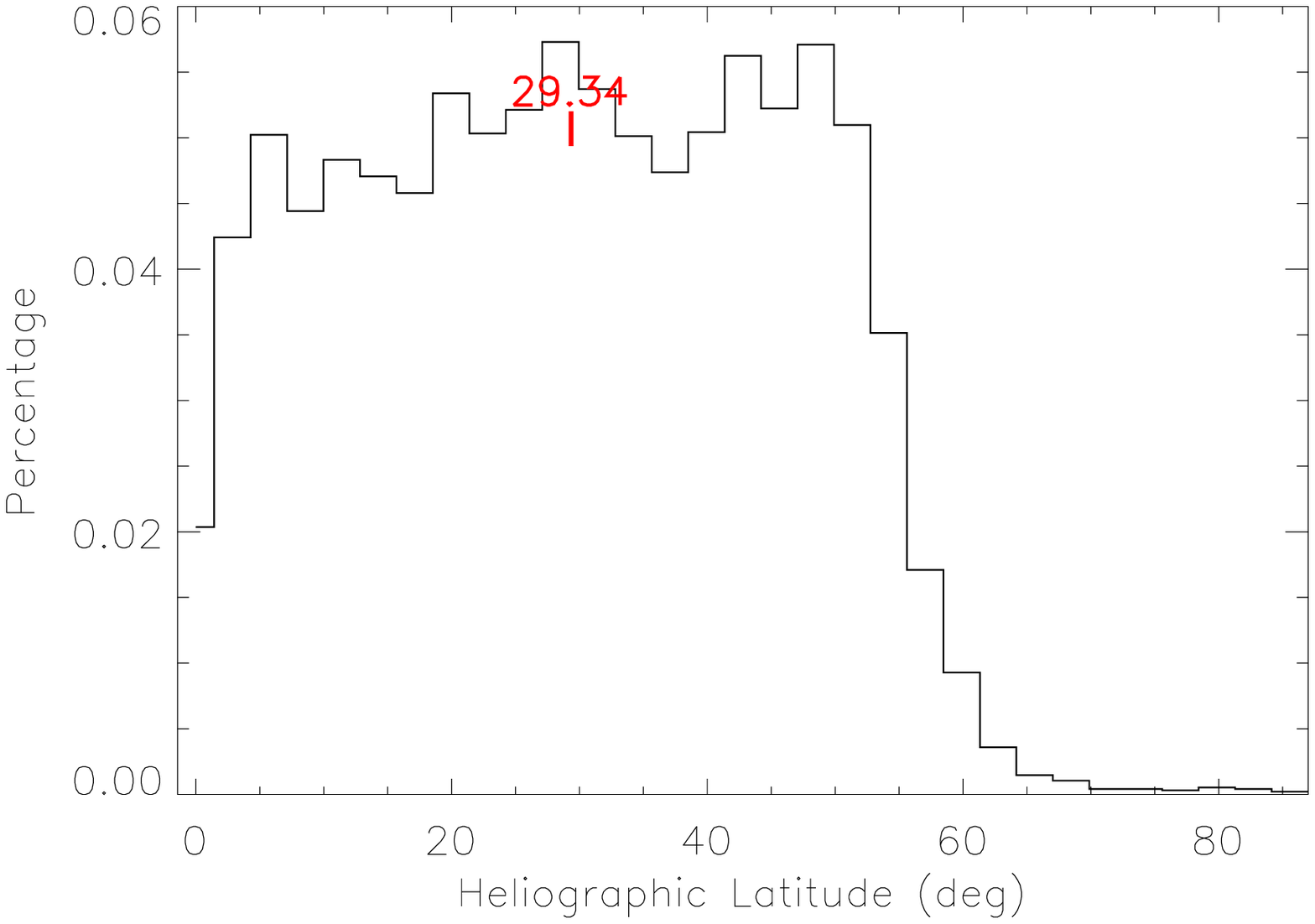}
  \includegraphics[width=0.49\hsize]{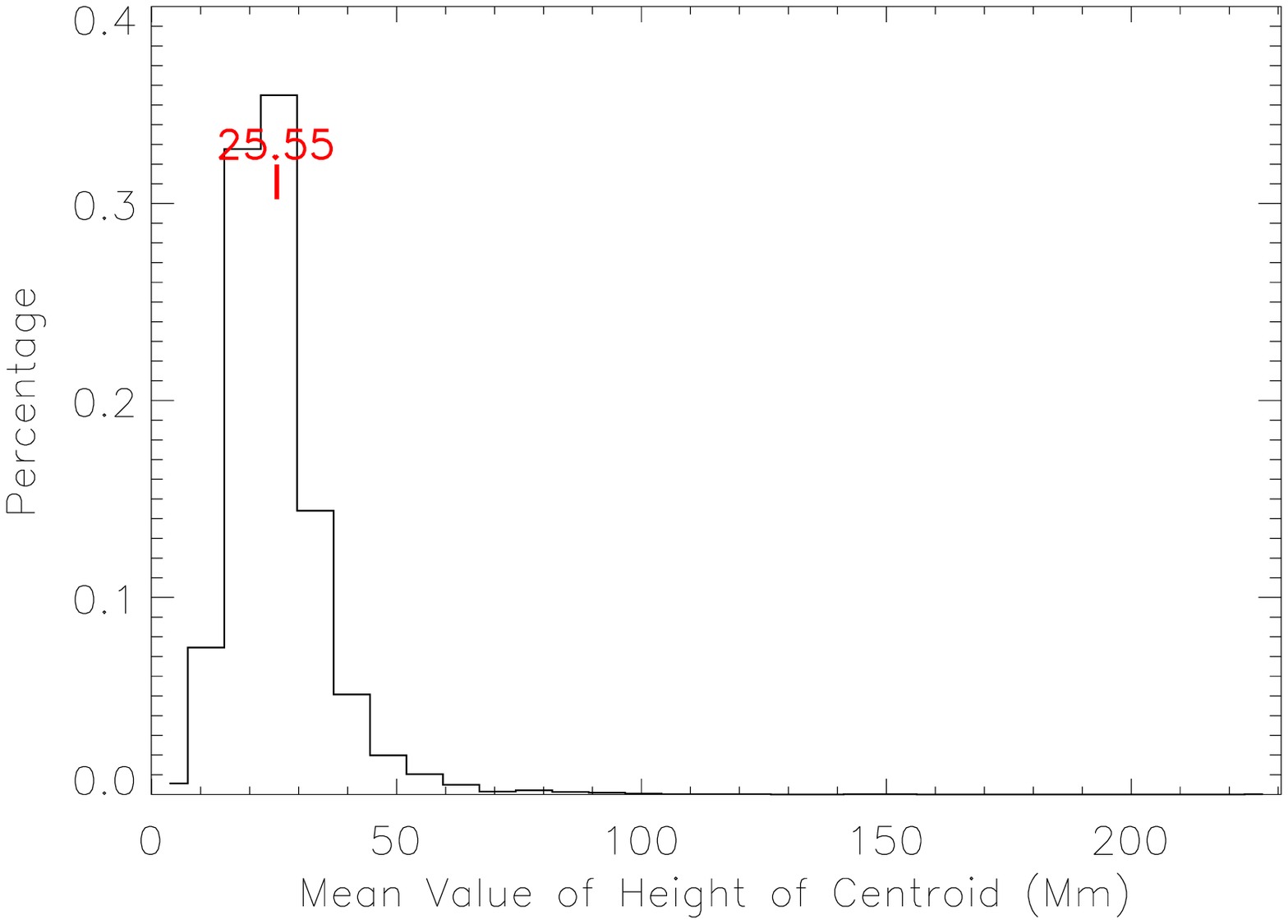}
  \includegraphics[width=0.49\hsize]{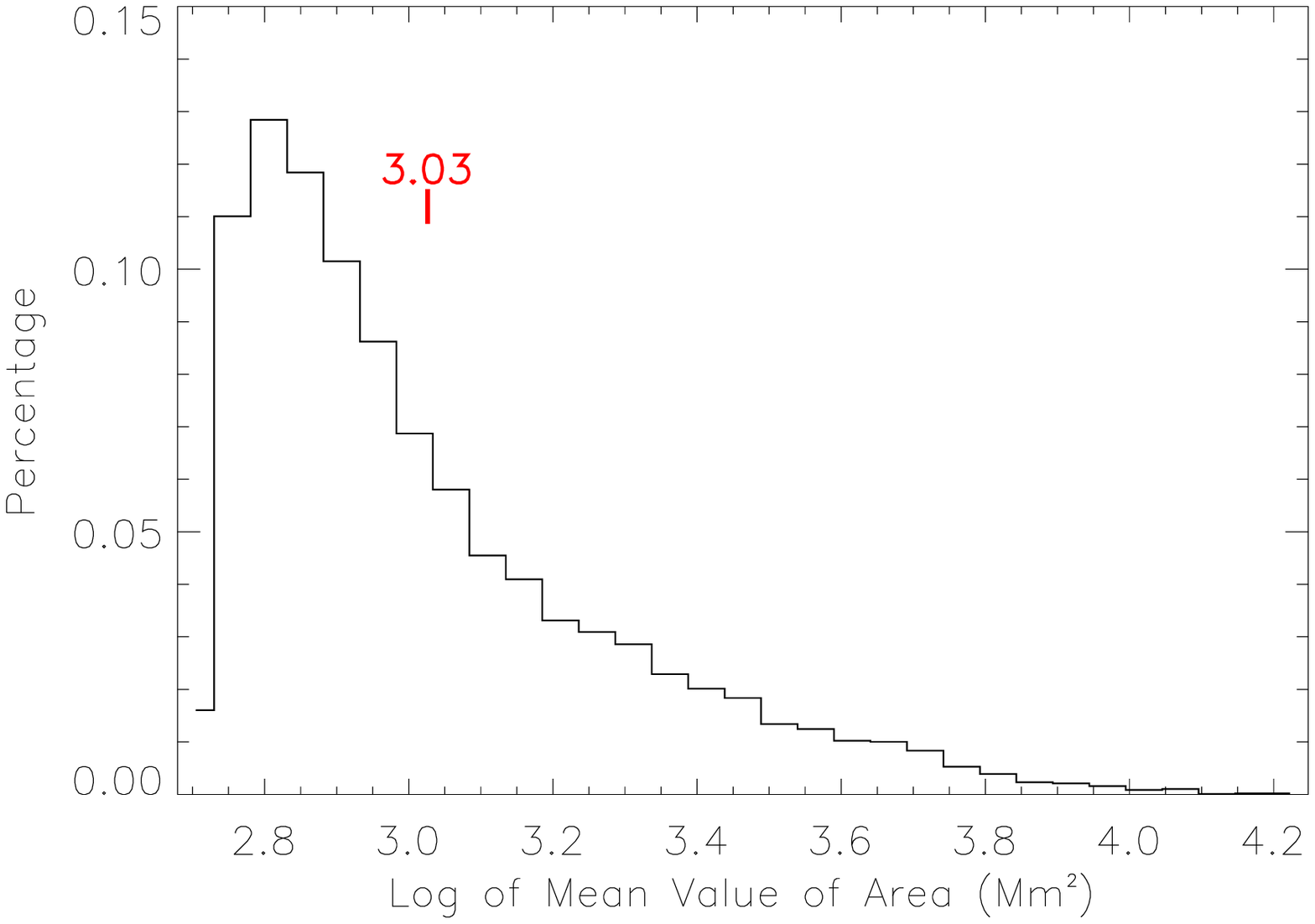}
  \includegraphics[width=0.49\hsize]{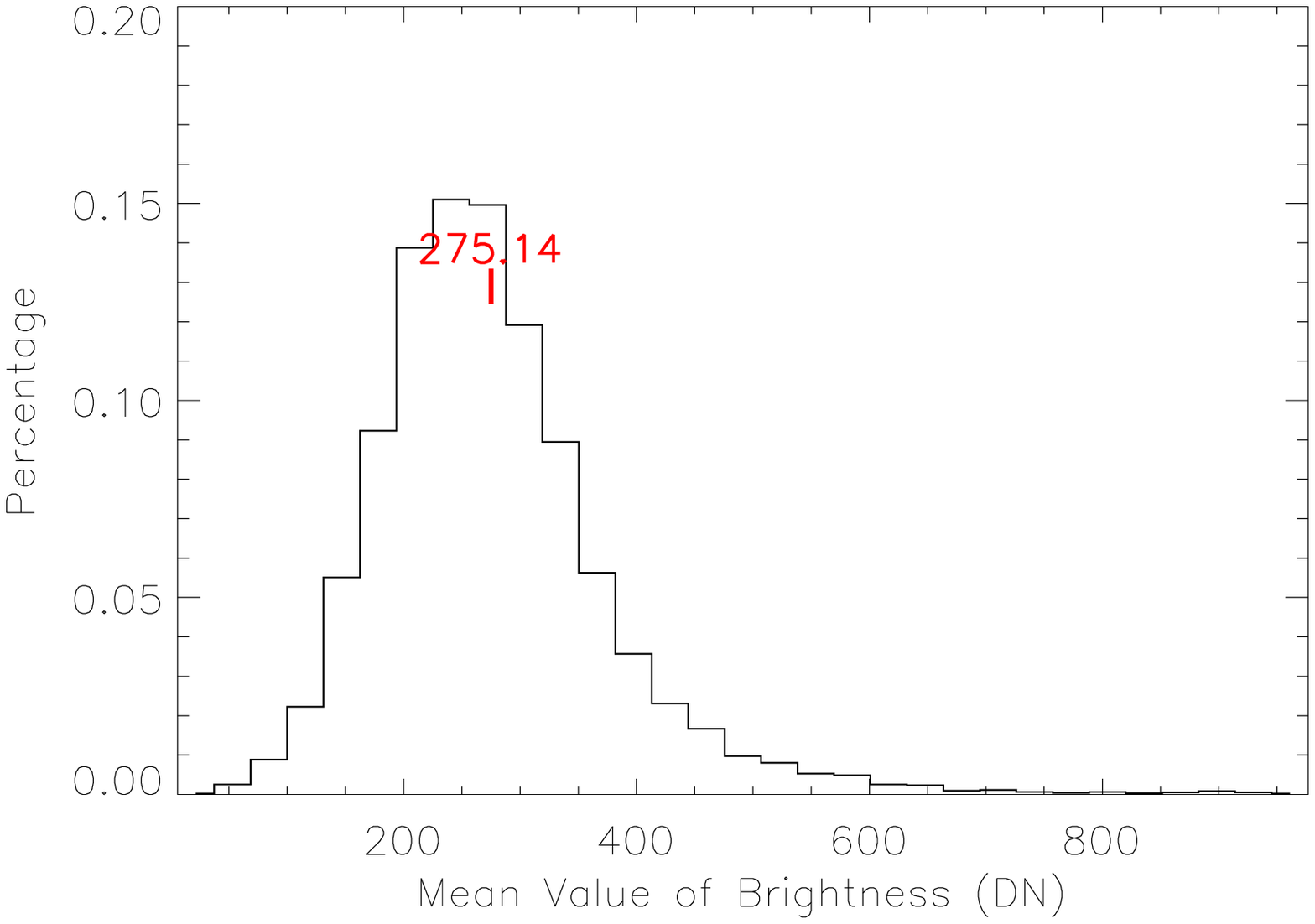}
  \caption{Histograms of static parameters: ({\it upper-left panel}) heliographic latitude at the first appearance,
mean values of ({\it upper-right panel}) height of centroid, ({\it lower-left panel}) area and
({\it lower-right panel}) brightness. The average values of these distribution are marked in the plots.}\label{fg_hist_static}
\end{figure*}

\begin{figure*}[tbh]
  \centering
  \includegraphics[width=0.49\hsize]{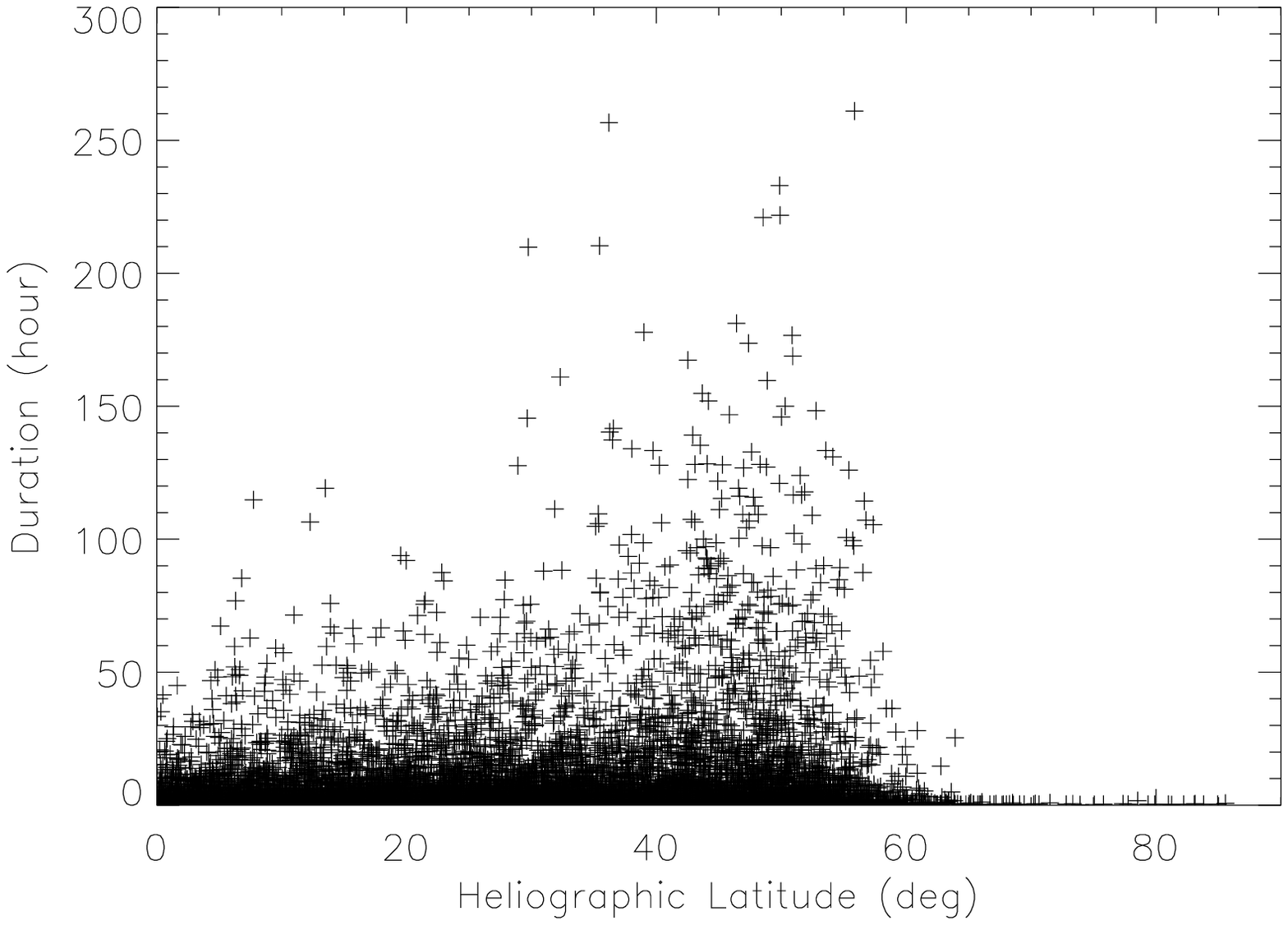}
  \includegraphics[width=0.49\hsize]{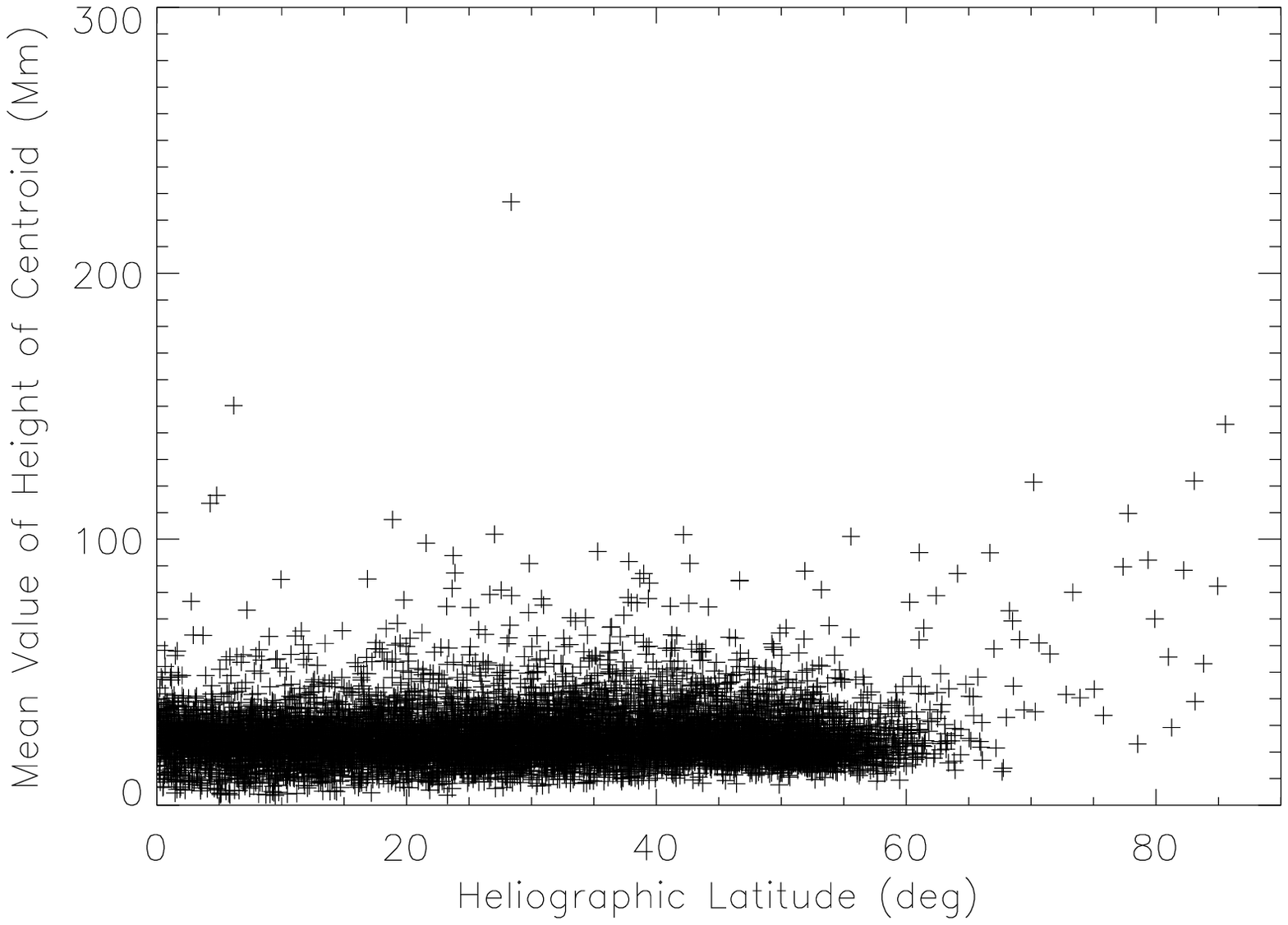}
  \includegraphics[width=0.49\hsize]{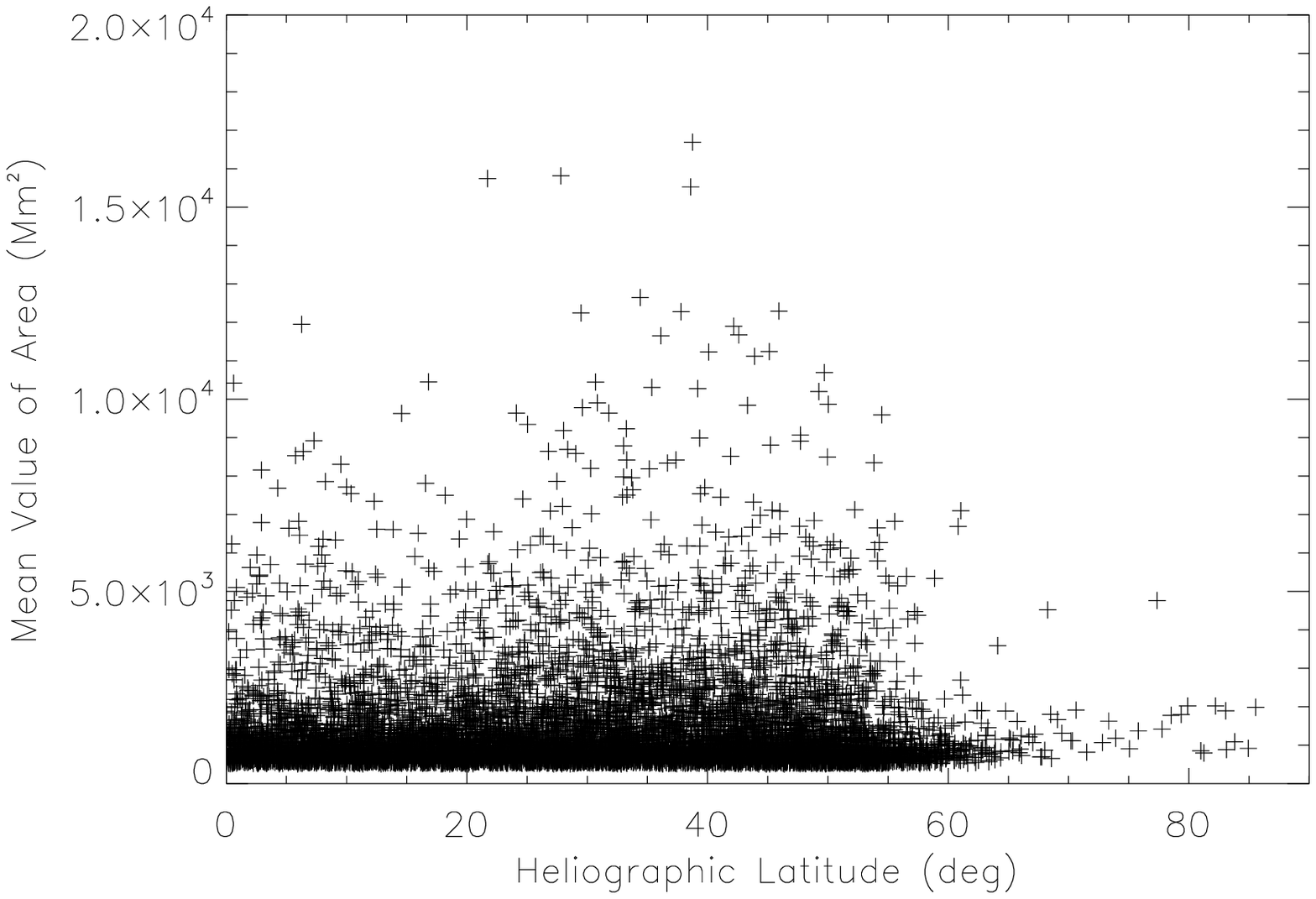}
  \includegraphics[width=0.49\hsize]{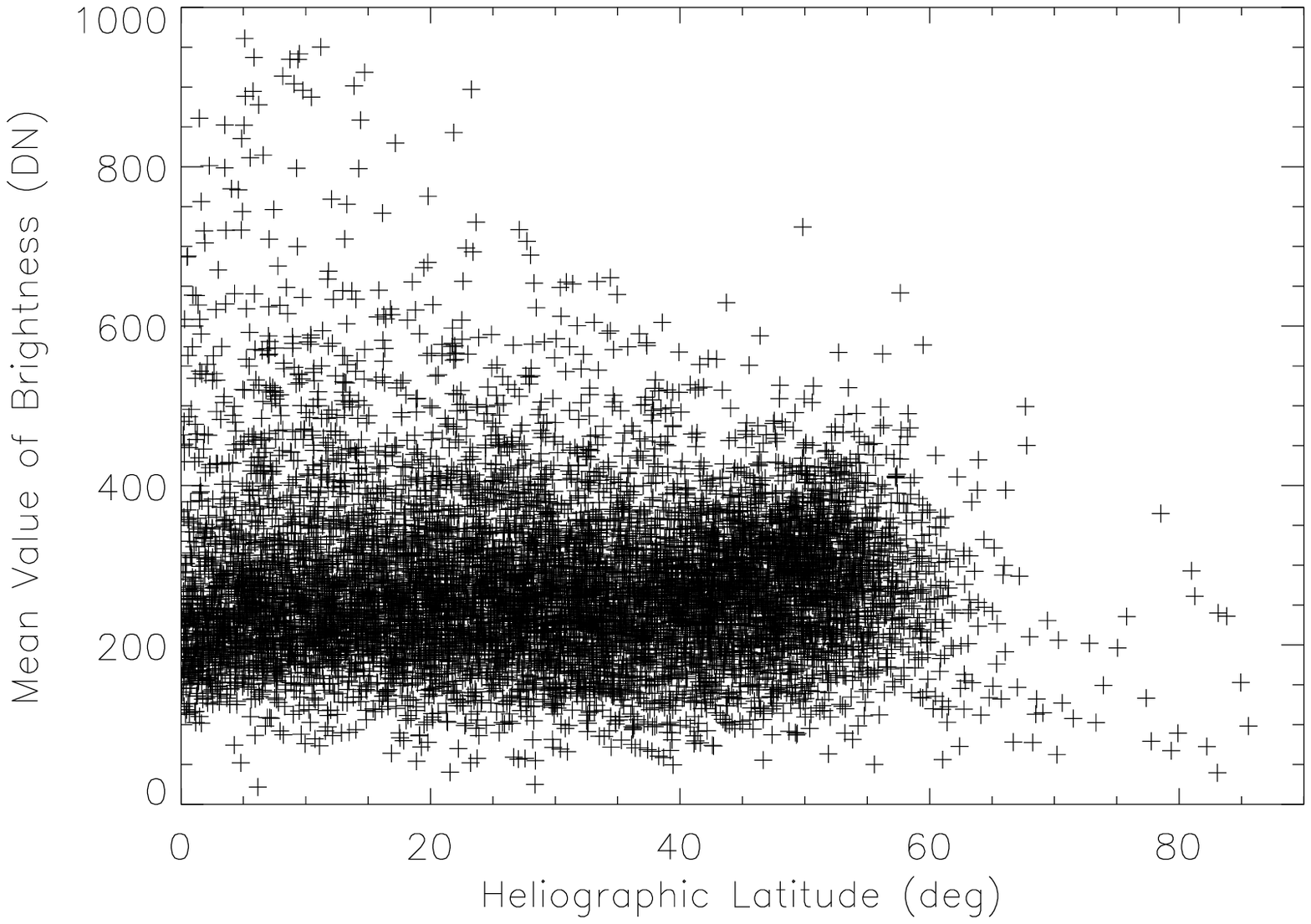}
  \caption{Correlations of heliographic latitude with ({\it upper-left panel}) duration,
mean values of ({\it upper-right panel}) the height of centroid,
({\it lower-left panel}) area, and ({\it lower-right}) brightness.}\label{fg_stat_lat}
\end{figure*}

First of all, static parameters of the 9477 well-tracked prominences are investigated. These parameters are (1) the heliographic latitude of the centroid of a prominence at the first detection, (2) the mean value of the height of the centroid, (3) the mean value of the area and (4) the mean value of the brightness of a prominence. The distribution of heliographic latitudes (folded at equator) shown in the upper-left panel of Figure \ref{fg_hist_static} suggests that 99\% prominences appear below 60 degrees. One may suspect that the lower count at high latitude may be caused by the perception that a prominence at high latitude is usually quiescent or a polar crown prominence and generally is long lived and extended. However, the scatter-plot of the latitude versus duration in the upper-left panel of Figure \ref{fg_stat_lat} indicates that this perception is wrong, as it clearly shows that the durations of the prominences above 60 degrees are short. The long-duration prominences appear around 30 to 60 degrees, which probably implies that long extended prominences arise there. Similar result can also be found in Figure \ref{fg_latitude} though there we include the poorly-tracked prominences. Moreover, from rescaled EUV 304~\AA\ images, one can clearly find that the regions above 60 degree are generally occupied by polar coronal holes. By checking the catalog and movies, we find that the detected `prominences' above 60 degrees are usually polar jets. Since their number  is small, the statistical results obtained below will not be affected by including these `false' prominences. 

The distribution of the heights of centroids suggests that about 82\% of prominences stay at around 26 Mm above the solar surface. There is no obvious dependence of the height on the latitude as shown by the scatter-plot in Figure \ref{fg_stat_lat}. The previous statistical study by \citet{Ananthakrishnan_1961} suggested that the heights of prominences vary between 15 and 150 Mm. \citet{Pettit_1932} gave a value of about 50 Mm, and \citet{Kim_etal_1988} showed that there is a peak at about 28 Mm in the distribution of heights. Theoretical work suggests that the height of a prominence depends on the gradient of ambient coronal magnetic field strength with height \citep[e.g.,][]{Filippov_Den_2000}. Our result is consistent with these studies.

The area and brightness can be usually used to evaluate if a prominence is a major one or not. Similarly, we use the mean values of them to show the distribution. The average projected area on the plane of sky of all prominences is about 1072 Mm$^2$. Nearly 60\% prominences have smaller area than the average value. Figure \ref{fg_stat_lat} suggests that the area is unrelated with the latitude. The brightness is recorded as digital number (DN) by the CCD camera. Its distribution is close to a Gaussian one, with an average value at around 275 DN. It can be as low as 20 DN or as high as 960 DN. The scatter-plot in Figure \ref{fg_stat_lat} shows a weak dependence of the brightness on the latitude. The low-latitude prominences may reach to a higher brightness than middle to high-latitude prominences.

\subsection{Dynamic Parameters}
\begin{figure*}[tbh]
  \centering
  \includegraphics[width=0.49\hsize]{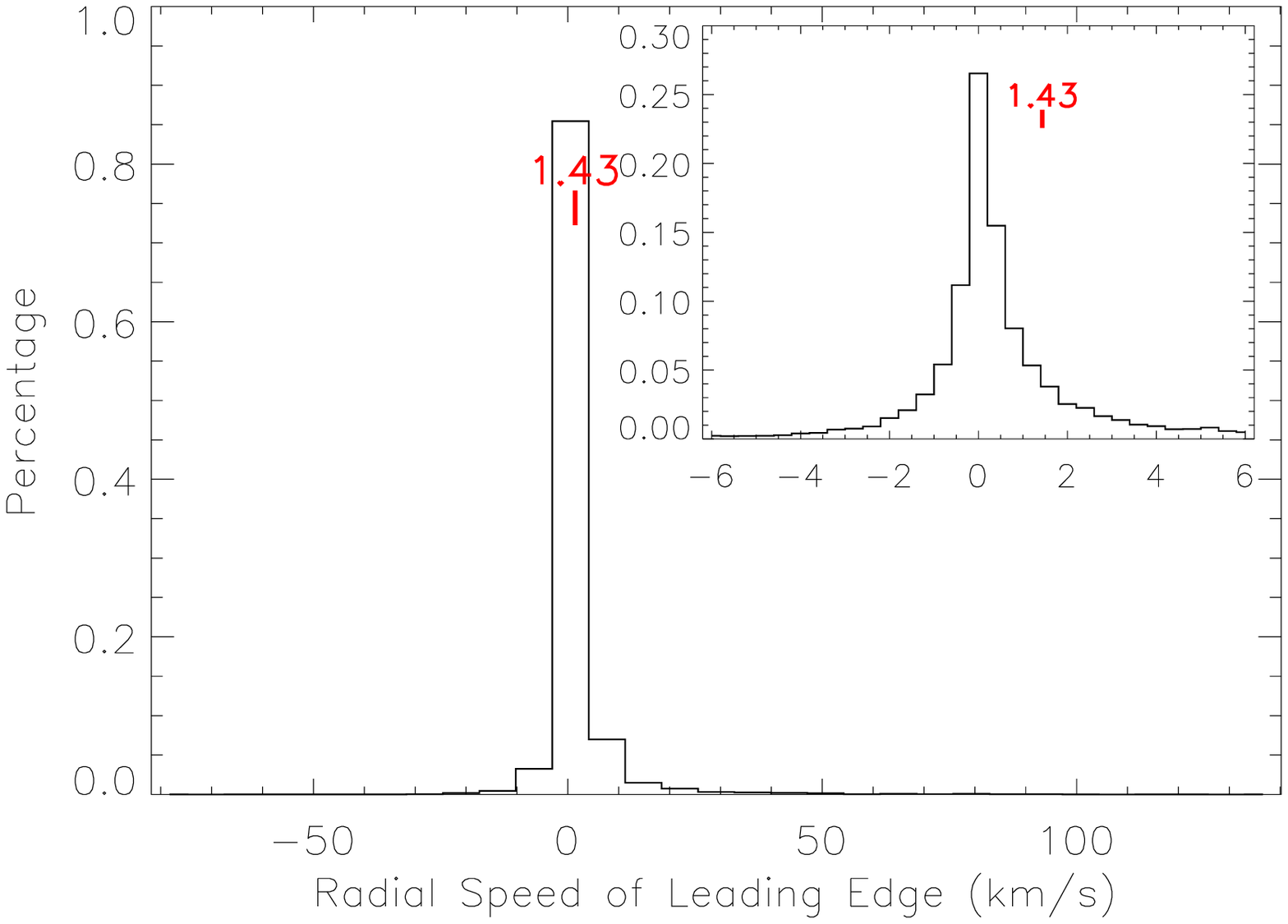}
  \includegraphics[width=0.49\hsize]{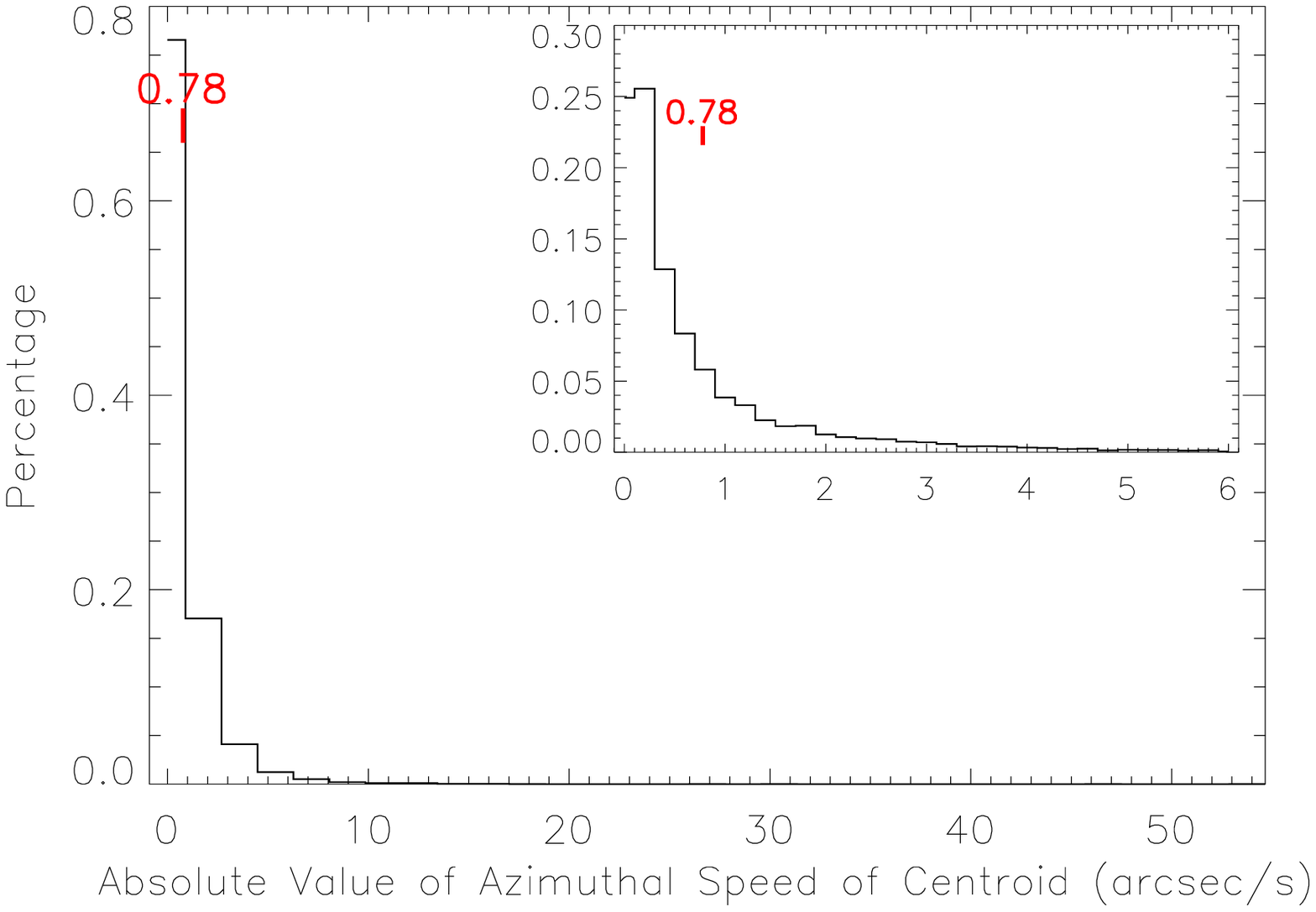}
  \includegraphics[width=0.49\hsize]{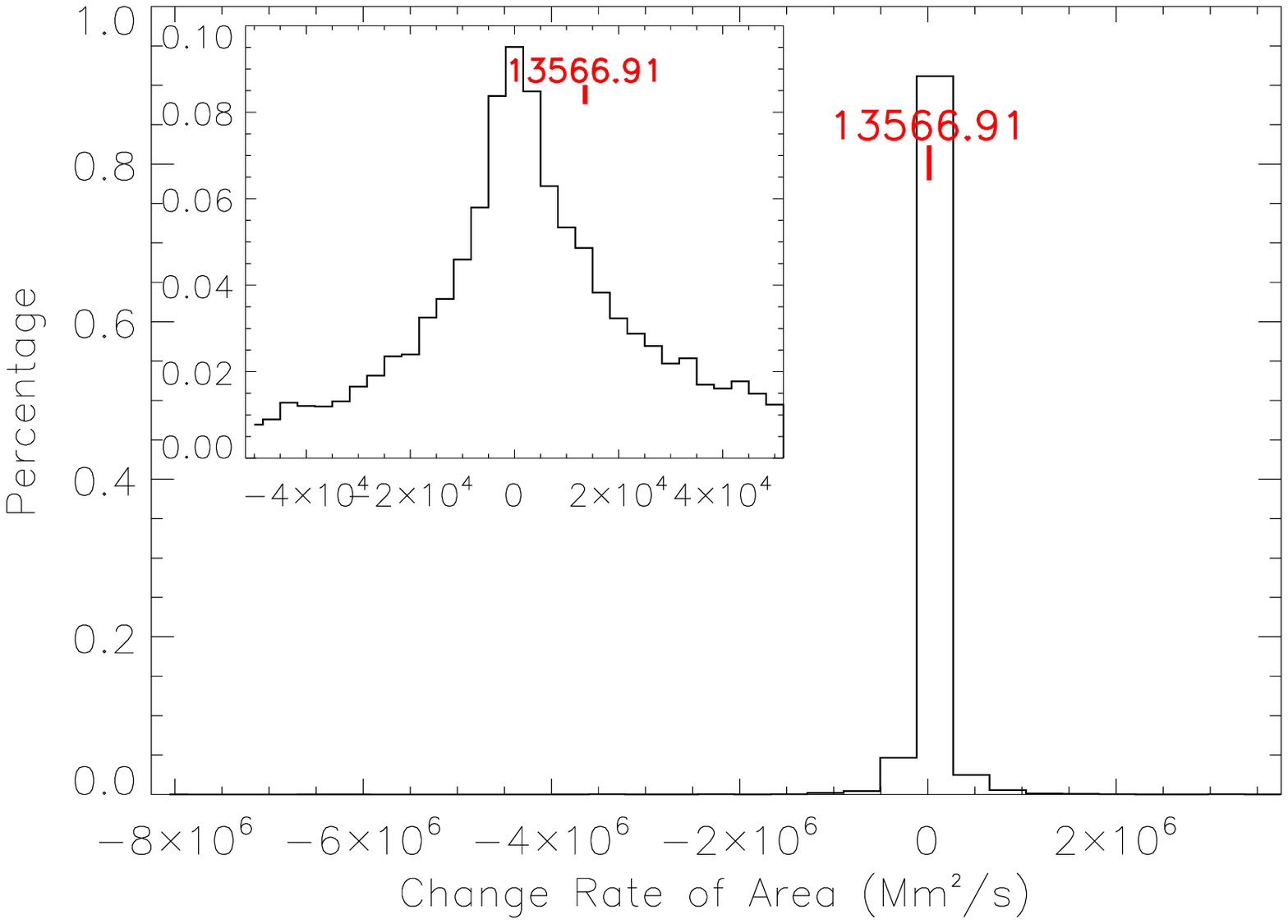}
  \includegraphics[width=0.49\hsize]{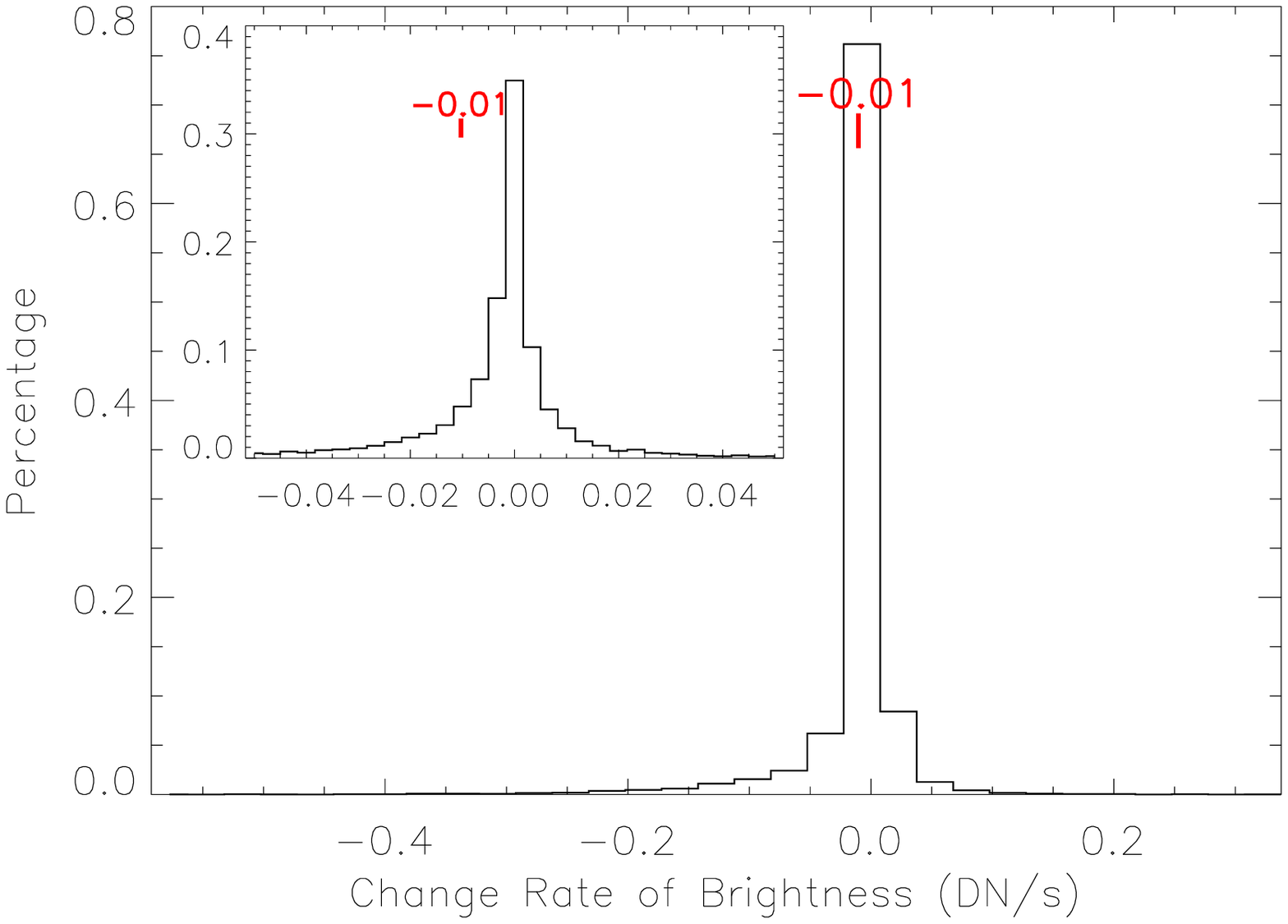}
  \caption{Histograms of dynamic parameters: ({\it upper-left panel}) radial speed of leading edge,
({\it upper-right panel}) azimuthal speed of centroid, ({\it lower-left panel}) change rate of area and
({\it lower-right panel}) brightness. The averaged values are marked in the plots.}\label{fg_hist_dynamic}
\end{figure*}

For dynamic properties, we investigate the velocities (both radial and azimuthal) and change rates of area and brightness of prominences. Here we use the leading edge rather than the centroid in the analysis of the radial speed. One also may use the radial speed of the centroid, but it will bring a large error in the case that a prominence splits into two parts, one erupting and the other staying on the Sun. The histograms in Figure \ref{fg_hist_dynamic} show the radial and azimuthal speeds. A speed of 10 arcsec/s corresponds to about 33 km/s at solar surface. More than 80\% of prominences have no obvious motion in either radial or azimuthal direction, and the value of zero is the most probable speed (as shown in the insets). A few prominences may move upward at about more than 100 km/s, and also there are 37 ($\approx0.4$\%) prominences having a radial speed $<-20$ km/s or an azimuthal speed $>10$ arcsec/s. The former could be easily understood that an erupting prominence may have a large outward speed, but the cause is not obvious for the latter.

We have mentioned in Sec.\ref{sec_performance} that the quality or precision of the fitting results largely depends on the number of measurements. More measurements can efficiently reduce errors, and therefore the fitting results will be more reliable. By checking the movies of the 37 prominences with the unusual speed, we find 34 prominences are detected in no more than 6 frames, and the speeds of most (not all of) these 34 prominences are not correctly reflected by the fitting. However, since we have a large sample in the statistics, such a small fraction of corrupted  events will not distort the overall picture shown in Figure \ref{fg_hist_dynamic}. On the other hand, it is also realized that there are indeed some prominences having a large downward or azimuthal speed. The large downward speed may either present a real motion or is just resulted from the shrink of the prominence. Any further analyses on such extreme events will be pursued in the future.

Similarly, for most prominences the change rates of area and brightness are quite small although the average value of the change rate of area is about $1.35\times10^4$ Mm$^2$/s. There are 106 (1.1\%) prominences with absolute value of change rate of area $>10^6$ Mm$^2$/s or brightness $>0.25$ DN/s. The movies reveal that some prominences do change this fast.

\subsection{Fading of Prominences}\label{sec_fading}

It is well known that prominences generally become dimmer as they rise. The reason could be the heating of prominence materials \citep[e.g.,][]{Mouradian_Martres_1986, Ofman_etal_1998, Hanaoka_Shinkawa_1999}, mass loss \citep[e.g.,][]{Rusin_Rybansky_1982} and/or expansion \citep[e.g.,][]{Bemporad_2009}. The first one belongs to thermal processes, and the other two are dynamic processes \citep[e.g.,][]{Mouradian_Martres_1986, Mouradian_etal_1995, Tandberg-Hanssen_1995}. Here we will look into this issue in a statistical way. In our parameters, we have the information of altitude, brightness and area of prominences, and have no direct information about the heating or mass of prominences. Thus it is impossible to make a comprehensive study of the causes of the prominence fading, but we can learn how significant the factor of the expansion is.

\begin{figure}[tbh]
  \centering
  \includegraphics[width=0.49\hsize]{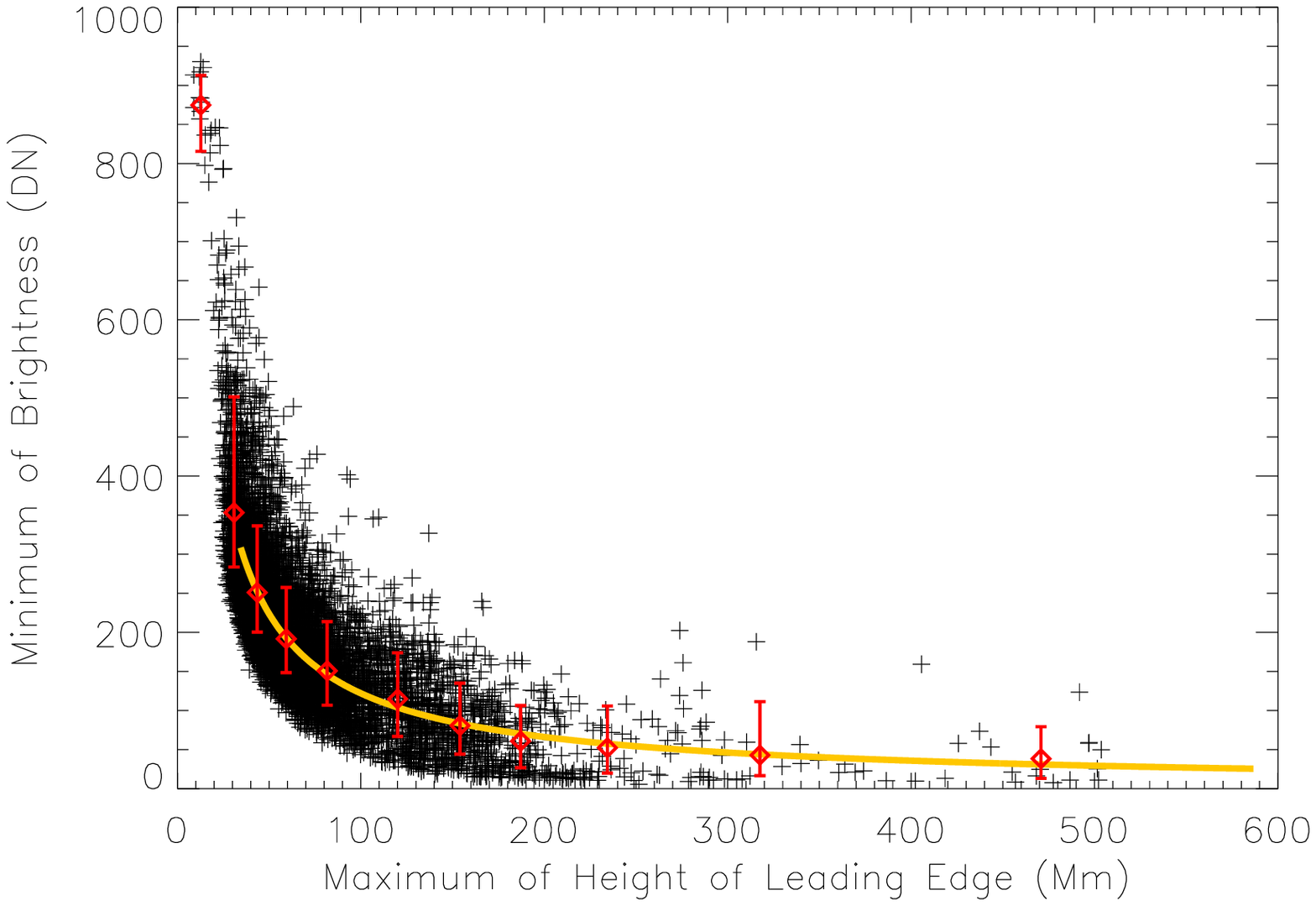}
  \includegraphics[width=0.49\hsize]{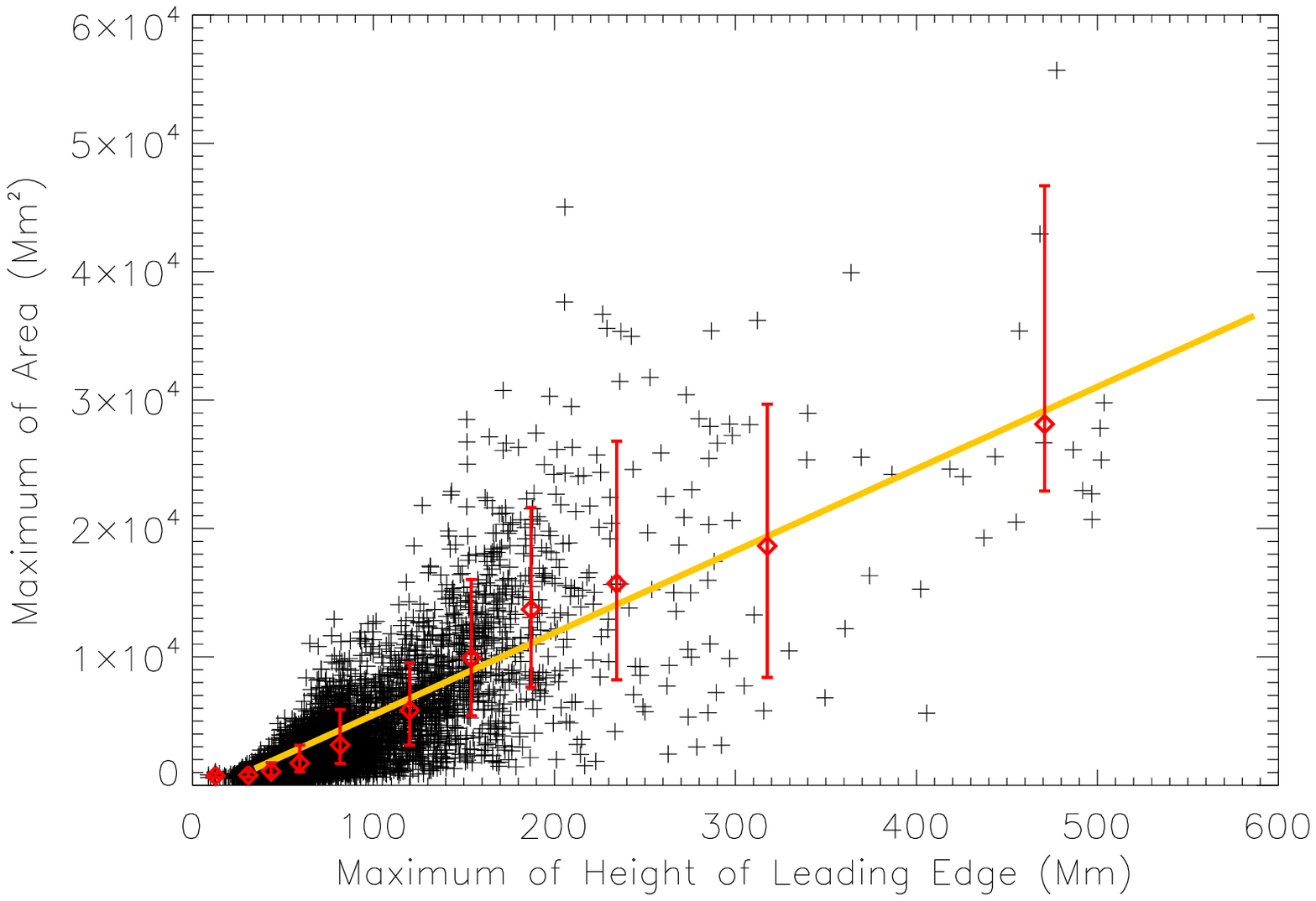}
  \includegraphics[width=0.49\hsize]{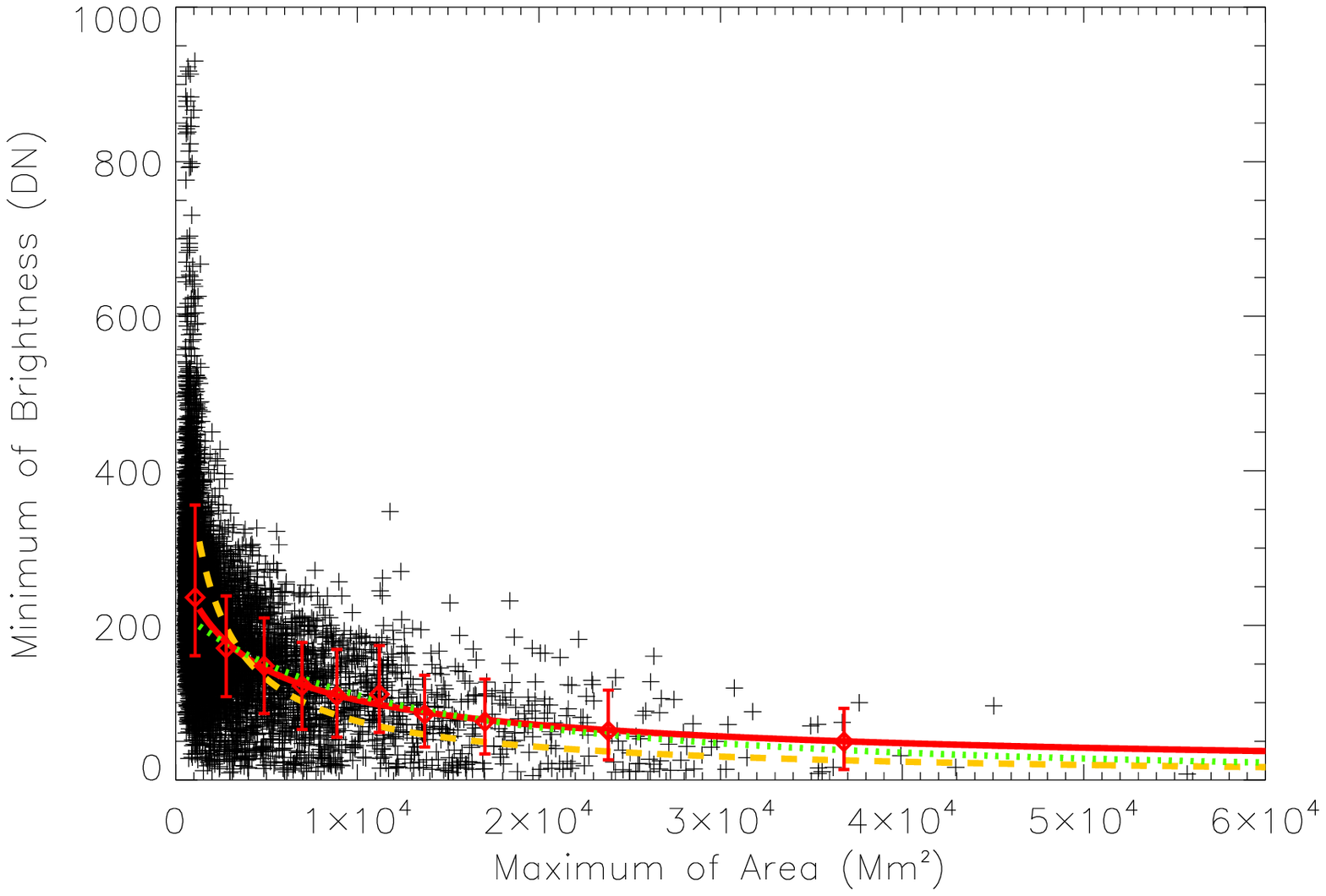}
  \caption{Scatter plots between minimum brightness, maximum area and maximum height
of leading edge.}\label{fg_stat_lbs}
\end{figure}

The first panel in Figure \ref{fg_stat_lbs} shows a strong anti-correlation between the brightness and the height of prominences. The higher altitude makes the prominence dimmer. The diamond symbols with error bars indicate the average values of brightness around certain altitude. These points are fitted to obtain the following empirical formula 
\begin{eqnarray}
F=7.47\times10^{3}(h+1.29)^{-0.891} \mathrm{\ DN} \label{eq_d-b}
\end{eqnarray}
where $h\geq35$ Mm is the height from solar surface in units of Mm. It describes the dependence of the brightness on the altitude. The points below 35 Mm are excluded in the fitting as they seem to follow another pattern.

The second panel exhibits an evident positive linear correlation between the area and height. It means a prominence at a higher altitude tends to be larger. This phenomenon supports the picture that, when a prominence rises or erupts, it expands as well. The expansion is probably caused by the weaker constraint of ambient atmosphere at a higher altitude. Similarly, we fit the data points marked by the diamond symbols, and get
\begin{eqnarray}
A=64h-906 \mathrm{\ Mm}^2  \label{eq_d-s}
\end{eqnarray}
The reversed correlations of Eq.\ref{eq_d-b} and \ref{eq_d-s} suggest that the expansion of prominences must be a cause of the prominence fading when they are rising or erupting.

Combining the two equations, we derive the relationship between the brightness and area (indicated by the dashed line in the last panel of Fig.\ref{fg_stat_lbs})
\begin{eqnarray}
F=3.04\times10^{5}(A+9.88\times10^2)^{-0.891} \mathrm{\ DN} \label{eq_b-s1}
\end{eqnarray}
where $A$ is in the units of Mm$^2$. Also we can directly fit the data in the last panel that leads to
\begin{eqnarray}
F=3.96\times10^{4}(A+2.42\times10^3)^{-0.631} \mathrm{\ DN} \label{eq_b-s2}
\end{eqnarray}
as shown by the solid line. Approximately, area is proportional to $V^{2/3}$ where $V$ is the volume of a prominence. If there is no mass loss or gain, the density of a prominence $\rho$ is inversely proportional to the volume, i.e., $\rho\propto V^{-1} \propto A^{-1.5}$. The brightness can be a proxy of the density of the plasma in the temperature window corresponding to EUV 304~\AA\ emission line. If there is no heating or cooling, the function $F=c_0(A+c_1)^{-1.5}$ is supposed to describe the relationship between the brightness and area. By fitting the data points, we obtain the dotted line given by
\begin{eqnarray}
F=4.76\times10^{8}(A+1.66\times10^4)^{-1.5} \mathrm{\ DN} \label{eq_b-s3}
\end{eqnarray}
It is found that the solid and dashed lines are close to the dotted one, which implies that, in a statistical point of view, the expansion is probably one of the major causes of the fading of prominences during their rise or eruption. Of course, this statistical conclusion could not be true for all individual cases. As revealed by, e.g., \citet{Ofman_etal_1998} and \citet{Hanaoka_Shinkawa_1999}, the heating process may play an important role in the disappearances of some prominences.

\section{Summary}
We have developed an automated system of catching and tracking solar limb prominences in EUV 304~\AA\ images. The system, called SLIPCAT, is able to generate (1) a catalog of solar limb prominences and (2) some characteristic parameters of each detected prominence, including the height, position angle, area, length, brightness and their first and second derivatives with respect to time. SLIPCAT is composed by five modules, (1) prominence candidate selection, (2) parameter extraction, (3) non-prominence feature removal, (4) prominence tracking, and (5) catalog generating. At present, an online catalog for STEREO-B/EUVI 304~\AA\ data has been generated (refer to \url{http://space.ustc.edu.cn/dreams/slipcat/}), and catalogs for STEREO-A/SECCHI/EUVI and SDO/AIA data are in preparation.

Based on the STEREO-B/EUVI 304~\AA\ data, SLIPCAT proved to perform well in detecting limb prominences. 
\begin{enumerate} 
\item It can distinguish real prominences from non-prominence features, e.g., active regions, without observations in other wavelengths by using the technique of linear discriminant analysis. The goodness of the overall classification is at about 86\% successful rate.

\item It detects as many as 9477 well-tracked prominences during 2007 April -- 2009 October, which means in average about 10 events per day. Compared to H$\alpha$ data, it is found that SLIPCAT is sensitive enough to recognize almost all prominences, even those invisible in H$\alpha$ images or very small ones.

\item Thanks to the high-cadence EUV 304~\AA\ data, SLIPCAT is able to provide the detailed evolution processes of prominences quantitatively without manual interventions. The upper-right panel of Figure \ref{fg_duration} implies that a well-tracked prominence is at least detected in 28 images in average; and the case in Figure \ref{fg_20081113_kin} shows that such high-cadence detection allows us to make a detailed analysis of  its evolution, including its eruption, oscillation, etc.
\end{enumerate}
However, not all the parameters extracted by SLIPCAT can precisely reveal the real behavior of prominences. The limitations have been addressed in the last paragraph of Sec.\ref{sec_performance}. Summarized here, they are that (1) the parameters characterize the properties of prominences during the period they are detected, not over the whole life-time, (2) they suffer from the projection effect, and (3) the speeds, change rates of area, length and brightness, which are derived from linear and quadratic fittings, may not be accurate.

By applying SLIPCAT to the STEREO-B/EUVI 304~\AA\ data from 2007 April to 2009 October, we obtain the following preliminary statistical results of solar limb prominences. 
\begin{enumerate} 
\item In average, there are about 10 prominences standing above the solar limb per day during the solar minimum. For most days, about 14 prominences are expected to be detected, and sometimes the number could be as large as 32 or as small as zero.

\item Most (99\%) prominences appear below latitude of 60 degrees, and the long extended prominences tend to arise between latitudes of 30 and 60 degrees.

\item Most (82\%) prominences have a height of about 26 Mm from the solar surface.

\item The projected area of a prominence on the plane-of-sky is about 1072 Mm$^2$ in average, and nearly 60\% of prominences have a smaller area.


\item Most prominences are quite stable during the period they are detected; no obvious change in position, area or brightness can be found.

\item Particularly, more than 80\% of prominences do not show obvious motion in either radial or azimuthal direction.  However, some prominences have an upward speed of more than 100 km/s, and a few prominences present a significant downward or azimuthal speed.

\item The brightness of prominences is anti-correlated with the height. The prominences at higher altitude look dimmer.

\item The area of prominences is positively correlated with the height. The prominences at higher altitude are generally larger.

\item From the statistical point of view, the expansion of prominences is probably one of the major causes of the fading of prominences during their rise or eruption.
\end{enumerate}

\paragraph{Acknowledgments.}
We acknowledge the use of the data from STEREO/SECCHI. This research is supported by grants from NSFC 40525014, 973 key project 2006CB806304, FANEDD 200530, and the fundamental research funds for the central universities.

\bibliographystyle{agufull}
\bibliography{../../ahareference}

\end{document}